\documentclass[11pt]{article}
\usepackage[margin=1in]{geometry}
\usepackage{cite}
\usepackage{amsmath,amssymb,amsfonts}
\usepackage{algorithmic}
\usepackage{graphicx}
\usepackage{rotating}
\usepackage{textcomp}
\usepackage{xcolor}
\usepackage{booktabs}
\usepackage{multirow}
\usepackage{url}
\usepackage{tikz}
\usetikzlibrary{shapes.geometric, arrows.meta, positioning, fit, backgrounds, calc}
\usepackage{pgfplots}
\pgfplotsset{compat=1.18}
\usepgfplotslibrary{groupplots}
\usepackage{array}
\usepackage{caption}
\newcolumntype{L}[1]{>{\raggedright\arraybackslash}p{#1}}
\usepackage[colorlinks=true,allcolors=blue!60!black]{hyperref}

\title{Landscape-Dependent Performance of Photonic Quantum Solvers in QUBO Feature Selection for Financial Risk Detection}

\author{
Nirvik Sahoo\\
\small Singapore Management University\\
\small 80 Stamford Road, Singapore 178902\\
\small\texttt{nirviks@smu.edu.sg}
\and
Paul R. Griffin\\
\small Singapore Management University\\
\small 80 Stamford Road, Singapore 178902\\
\small\texttt{paulgriffin@smu.edu.sg}
}
\date{}

\begin{document}
\maketitle


\begin{abstract}
Feature selection for imbalanced classification tasks such as credit card fraud and consumer default detection requires balancing predictive relevance, inter-feature redundancy, and computational feasibility. We benchmark three computing paradigms, classical branch and bound optimization (Gurobi), photonic entropic computing (QCI Dirac-3), and photonic boson sampling (Piquasso\footnote{Provided by QuiX, \url{https://www.quixquantum.com/}.}), across thirteen feature-selection methods on two datasets: ULB Credit Card Fraud (30 features) and AmEx consumer default (159 features). Each method class is routed to the solver best matched to its mathematical structure and native capabilities: Gurobi solves standard and graph-theoretic QUBOs to certified optimality with respect to the formulated objective, Dirac-3 supports higher-order and integer-weighted objectives, and Piquasso samples feature subsets via permanent-weighted sampling. On the ULB dataset, Dirac-3 MI-Spearman using only 13 of 30 features reaches a mean F1 of $0.873 \pm 0.023$ over five runs (best run $0.896$), and Piquasso is the best method at $k{=}5$. In contrast, the AmEx dataset shows no comparable compression advantage, with performance improving steadily as more features are retained and all three paradigms approaching F1 $\approx 0.80$ only at large feature budgets ($k{=}150$). The AmEx dataset also highlights a disconnect between optimization objective value and downstream predictive performance: for the distance-correlation method at $k{=}25$, Gurobi’s globally optimal solution achieves F1 $=0.422$, while Dirac-3 achieves a mean F1 of $0.746$ on the same method. Using Jaccard similarity to compare selected feature subsets, we find that solver and method choices often produce different selections, but the performance impact of those differences is strongly dataset-dependent: at matched feature budgets, F1 varies about ten times more across methods on ULB than on AmEx. Overall, our results show that effective feature selection requires balancing predictive relevance and redundancy against the computational capabilities of the solver: specialized solvers matter most in compact, largely uncorrelated feature spaces, and less in larger, more redundant ones.

\end{abstract}


\section{Introduction}
\label{sec:intro}
Feature selection for imbalanced financial data, such as credit card fraud or consumer default, has to balance three things: how predictive each feature is, how much features overlap with each other, and how expensive the search is. In the standard ULB benchmark dataset \cite{ulb_kaggle}, confirmed fraud accounts for only 0.17\% of 284,807 transactions, making accuracy useless as a metric and demanding measures that reflect performance on the minority class directly. Feature selection over the 30-dimensional input space, consisting of Time, Amount, and PCA-anonymized components V1--V28, is essential for three operationally relevant reasons: faster real-time scoring, reduced overfitting to noise via a smaller and less-redundant input space, and simpler model governance and auditability. Because V1--V28 are anonymized PCA components, no analyst can judge feature relevance from its name; selection must be done entirely on mathematical grounds.

The AmEx consumer default dataset \cite{howard2022amex} presents a structurally different feature-selection problem at more than five times the dimensionality: 159 candidate features (after pre-processing) spanning payment, balance, spend, delinquency, and risk categories, evaluated over a much larger sweep of feature budgets. ULB’s compact PCA feature space permits aggressive compression, but because its components are largely uncorrelated, performance depends strongly on retaining the most informative features. AmEx has more than five times as many features, with predictive signal spread across many partly redundant features (Section~\ref{sec:dependence}), and it is unclear whether the solver and method choices, or the compression-performance trade-off, that work on ULB also hold there. Answering that question is one of the main goals of this paper.

The dominant formulation of feature selection as combinatorial optimization is the Quadratic Unconstrained Binary Optimization (QUBO) framework \cite{kochenberger2014unconstrained}, in which each candidate feature is assigned a binary decision variable and the objective trades pairwise relevance against pairwise redundancy. The mRMR criterion \cite{peng2005feature} is a widely used instantiation. The published reference benchmark \cite{samanta2025aaai} applies QUBO-based methods to both datasets using D-Wave's Advantage hybrid quantum annealer: on ULB, it reports a best F1 of 0.87356 (MI-Spearman, $k{=}25$, classical method) and a best Gini of 97.678\% (MI-Pearson, $k{=}13$); on AmEx, it reports a best F1 of 0.68827 and a best Gini of 85.07098\% (both MIQUBO, $k{=}150$). We cite these values for reference only: they were obtained with a different solver and, on AmEx, a different feature working set (Section~\ref{sec:dataset}), so our comparisons are made between our own three solver paradigms and against an all-features model trained on the same data.

This paper makes four contributions. First, we introduce a solver-aware design principle: each method class is matched to the solver that can handle its mathematical structure exactly, rather than approximately or not at all. Standard and graph-theoretic QUBO problems go to Gurobi and Dirac-3; additionally, hardware-native methods (higher-order objectives and integer-weighted variables) go to Dirac-3; permanent-weighted sampling goes to Piquasso. Second, we provide a comprehensive empirical sweep across both datasets ($k \in \{5, 10, 13, 15, 20, 25\}$ on ULB and $k \in \{25, 50, 75, 100, 125, 150\}$ on AmEx), spanning 13 methods and three solver paradigms, tracing the compression-performance frontier for each method class on each dataset. Third, we measure how much the selected feature subsets overlap, using Jaccard similarity. We compare subsets across methods within each of the independent Dirac-3 runs, and between Gurobi and Dirac-3. Every value is compared with the overlap expected between random subsets of the same size, so that genuine agreement can be separated from overlap that arises by chance. Fourth, we run the same benchmark on two datasets that differ in size and structure. This lets us test whether the conclusions from the smaller dataset, about which methods suit which solvers and whether aggressive feature compression pays off, still hold on the larger one.

The paper is organized as follows. Section~\ref{sec:related} reviews related work. Section~\ref{sec:dataset} describes the datasets and evaluation metrics. Section~\ref{sec:methods} presents the three solver paradigms and their associated methods. Section~\ref{sec:experiments} details the experimental setup. Section~\ref{sec:results} reports and interprets results. Section~\ref{sec:diversity} analyzes subset diversity. Section~\ref{sec:discussion} discusses broader findings. Section~\ref{sec:conclusion} concludes.

\section{Related Work}
\label{sec:related}
\subsection{Information-Theoretic and QUBO Feature Selection}
All the QUBO-based methods in this paper build on the relevance-redundancy trade-off formalized by mRMR \cite{peng2005feature}, which favors features that are informative about the target and not redundant with each other. Brown et al.~\cite{brown2012conditional} unified a broad family of mutual-information-based criteria under a conditional likelihood framework, showing that many classical selectors are special cases of a common objective. Li et al.~\cite{li2017feature} survey feature selection from a data perspective, organizing methods by the kind of data they handle and by their underlying criteria.

Quantum annealing for feature selection was applied to credit scoring by Milne et al.~\cite{milne2017optimal} and to recommender systems by Nembrini et al.~\cite{nembrini2021feature}, both using D-Wave hardware. M\"ucke et al.~\cite{mucke2023feature} formulate feature selection as a QUBO with a tunable relevance-redundancy weight and solve it on both quantum annealing and gate-based hardware. The direct predecessor of the present work, Samanta et al.~\cite{samanta2025aaai}, formulates six QUBO variants and evaluates them on both the ULB fraud dataset and the AmEx consumer default dataset with the D-Wave hybrid solver, and reports F1 and Gini values on each dataset that we cite as reference points.

\subsection{Graph-Theoretic Feature Selection}
Graph structure can guide feature selection beyond pairwise scores. Zhao and Liu \cite{zhao2007spectral} introduce spectral feature selection, which ranks features by how consistent they are with the structure of a sample-similarity graph, measured through its Laplacian. We adapt this idea to a graph whose nodes are features and whose edges are pairwise correlations, using spectral leverage as a relevance bonus. Community detection offers a second route: grouping correlated features into communities, here with the Louvain method \cite{blondel2008fast}, lets a selection be required to draw from several groups rather than one. Graph attention networks \cite{velickovic2018graph}, which learn how much weight each node assigns to its neighbours, motivate our attention-weighted redundancy penalty.

\subsection{Higher-Order and Hardware-Native Quantum Methods}
Fleuret \cite{fleuret2004fast} used conditional mutual information as a filter criterion for fast binary feature selection; our HUBO method uses conditional mutual information between features to build cubic redundancy terms. Boros and Hammer \cite{boros2002pseudo} survey pseudo-Boolean optimization, including the reduction of higher-order terms to quadratic ones. That reduction requires auxiliary variables, which enlarges the problem, so hardware that handles cubic terms natively avoids this overhead. The native capabilities of QCI Dirac-3 (multi-level qudit variables and cubic objective terms) are described in \cite{qci2024dirac3,nguyen2025entropy}.

\subsection{Boson Sampling}
Boson sampling was introduced by Aaronson and Arkhipov \cite{aaronson2011computational}: single photons sent through a linear-optical interferometer produce output patterns whose probabilities are given by squared matrix permanents, and sampling from this distribution is believed to be hard for classical computers under plausible complexity-theoretic assumptions. Bromley et al.~\cite{bromley2020applications} survey near-term applications of photonic devices to graph and combinatorial problems, most of which use the Gaussian variant of boson sampling. We instead use standard boson sampling with single-photon inputs, so that sampled feature subsets are weighted by permanents. The sampler is implemented in the Piquasso framework \cite{Piquasso2024processor}, which provides simulators for both standard and Gaussian boson sampling.

\section{Dataset and Evaluation}
\label{sec:dataset}
\subsection{ULB Credit Card Fraud Dataset}
The ULB dataset \cite{ulb_kaggle} contains 284,807 European credit card transactions over a 48-hour window in September 2013. Of these, 492 (0.17\%) are confirmed fraud. The 30 features consist of Time (elapsed seconds from the first transaction), Amount, and V1--V28, the top 28 components of a PCA transformation of the original transaction features. Because the PCA components carry no human-readable meaning, feature selection cannot rely on domain judgment and must be driven entirely by statistical relationships with the fraud label and with each other.

Table~\ref{tab:dataset-ulb} summarizes the dataset characteristics and their implications for this benchmark.
\begin{table}[htbp]
\caption{ULB Dataset Characteristics}
\label{tab:dataset-ulb}
\centering
\footnotesize
\begin{tabular}{@{}L{0.24\columnwidth}L{0.24\columnwidth}L{0.42\columnwidth}@{}}
\toprule
\textbf{Property} & \textbf{Value} & \textbf{Implication} \\
\midrule
Total transactions & 284,807 & Minority-class stats are stable \\
Confirmed fraud & 492 (0.17\%) & Accuracy is uninformative \\
Features & 30 (Time, Amount, V1--V28) & Fixed, moderate-dimensional pool \\
Feature transparency & V1--V28 are PCA-anonymized & Manual audit is not viable \\
Time span & 2 days (Sept.\ 2013) & Short window; limited temporal diversity \\
\bottomrule
\end{tabular}
\end{table}

\subsection{AmEx Consumer Default Dataset}
The AmEx dataset \cite{howard2022amex} contains monthly consumer credit default profiles, with the target variable indicating whether a customer subsequently defaults. Candidate features span five semantically named categories: Payment, Balance, Spend, Delinquency, and Risk. After removing non-numeric, zero-variance, and fully-missing columns, 159 features remain for QUBO and graph-based construction. Unlike ULB's anonymized PCA components, AmEx's feature names carry categorical meaning, but individual features are numerous and partly redundant (Section~\ref{sec:dependence}), so redundancy reduction remains a statistical problem.

Table~\ref{tab:dataset-amex} summarizes the dataset characteristics and their implications for this benchmark, contrasted against ULB.
\begin{table}[htbp]
\caption{AmEx Dataset Characteristics}
\label{tab:dataset-amex}
\centering
\footnotesize
\begin{tabular}{@{}L{0.24\columnwidth}L{0.24\columnwidth}L{0.42\columnwidth}@{}}
\toprule
\textbf{Property} & \textbf{Value} & \textbf{Implication} \\
\midrule
Features (post-filter) & 159 & Over 5$\times$ ULB's dimensionality \\
Feature categories & 5 (Payment, Balance, Spend, Delinquency, Risk) & Semantically named but still statistically redundant \\
Feature transparency & Category-level names, no per-feature semantics & Partial audit only \\
\bottomrule
\end{tabular}
\end{table}

\subsection{Feature Dependence Structure}
\label{sec:dependence}

The two datasets differ in how the predictive signal is distributed. Table~\ref{tab:dependence} summarizes both.

\begin{table}[htbp]
\caption{Feature dependence structure. $|r|$: absolute Pearson correlation; $|\rho|$: absolute Spearman correlation; $|r_{y}|$: absolute correlation with the label. AmEx values are from a 2{,}000-customer sample (146 features after preprocessing).}
\label{tab:dependence}
\centering
\footnotesize
\begin{tabular}{lcc}
\toprule
\textbf{Statistic} & \textbf{ULB} & \textbf{AmEx} \\
\midrule
Mean pairwise $|r|$, all features          & 0.013 & 0.089 \\
Mean pairwise $|\rho|$, all      & 0.072 & 0.074 \\
Max pairwise $|\rho|$, V1--V28             & 0.68 (V21, V22) & -- \\
Mean $|r|$ within / between categories     & -- & 0.124 / 0.075 \\
Near-duplicate pairs ($|r|>0.95$)          & 0 & 11 \\
Features with $|r_{y}|>0.3$                & 2 & 29 \\
Features with $|r_{y}|>0.2$                & 4 & 48 \\
Largest $|r_{y}|$                          & 0.33 (V17) & 0.64 (P\_2) \\
Median $|r_{y}|$                           & 0.04 & 0.15 \\
\bottomrule
\end{tabular}
\\[6pt]
\begin{tabular}{lccccc}
\toprule
\multicolumn{6}{l}{\textit{AmEx: mean $|r|$ within and between categories}} \\
 & \textbf{Payment} & \textbf{Balance} & \textbf{Spend} & \textbf{Delinq.} & \textbf{Risk} \\
\midrule
Payment     & 0.275 & 0.148 & 0.090 & 0.138 & 0.125 \\
Balance     &       & 0.213 & 0.082 & 0.099 & 0.058 \\
Spend       &       &       & 0.114 & 0.057 & 0.038 \\
Delinquency &       &       &       & 0.102 & 0.061 \\
Risk        &       &       &       &       & 0.107 \\
\bottomrule
\end{tabular}
\end{table}

\textbf{ULB.} The principal components V1--V28 are linearly uncorrelated by construction (pairwise Pearson $|r|=0$ to three decimals); only Time and Amount correlate with some components (e.g.\ Amount with V2, $|r|=0.53$; Time with V3, $|r|=0.42$). The components are not independent, however: their mean pairwise Spearman correlation is 0.072, with pairs such as V21--V22 reaching 0.68. Rank-based and mutual-information redundancy measures can therefore detect structure that Pearson correlation cannot. The fraud signal is concentrated in a few components: only V17 ($|r_y|=0.33$) and V14 (0.30) exceed $|r_y|=0.3$, followed by V12 (0.26) and V10 (0.22), and the median feature has $|r_y|=0.04$. A small subset either contains these few components or misses most of the signal.

\textbf{AmEx.} Features are moderately correlated overall (mean $|r|=0.089$, about seven times ULB's), with somewhat higher correlation within categories than between them (0.124 vs.\ 0.075). The category structure is not a set of cleanly separated blocks: Payment correlates more strongly with Balance and Delinquency than Spend, Delinquency or Risk features correlate with each other. Redundancy is concentrated in a small number of near-duplicate pairs (11 with $|r|>0.95$, e.g.\ D\_139 and D\_143). Above all, the default signal is spread across many features: 29 features have $|r_y|>0.3$ and 48 have $|r_y|>0.2$, so many different subsets can capture most of the available signal.

\subsection{Evaluation Metrics}
Two metrics are reported for every (method, $k$, solver) configuration on both datasets. They capture distinct aspects of model quality and are not always maximized by the same feature subset; this distinction is central to interpreting the results.

\textbf{F1 Score} is the harmonic mean of precision and recall:
\begin{equation}
    \text{F1} = \frac{2 \cdot \text{Precision} \cdot \text{Recall}}{\text{Precision} + \text{Recall}}.
    \label{eq:f1}
\end{equation}
On both datasets, the target class is a minority (0.17\% for ULB fraud; a smaller but non-trivial share of accounts for AmEx default), so a classifier that always predicts the majority class achieves misleadingly high accuracy while catching none of the target class. F1 is therefore the primary metric for evaluating precision-recall balance on the minority class in both settings.

\textbf{Gini Coefficient} measures ranking quality independently of any decision threshold. It relates to the area under the ROC curve by $\text{Gini} = 2 \cdot \text{AUC} - 1$ and is the standard performance metric in credit-risk modelling. A method can rank transactions or accounts by risk very accurately (high Gini) while failing to set the optimal decision cutoff (lower F1), or vice versa. Any production deployment should specify which metric is primary before selecting a method.

\subsection{Reference Results from Prior Work}
The reference \cite{samanta2025aaai} reports, on ULB:
\begin{itemize}
    \item Best F1: 0.87356 (MI-Spearman, $k{=}25$, Classical solver)
    \item Best Gini: 97.678\% (MI-Pearson, $k{=}13$, D-Wave hybrid solver)
\end{itemize}
and, on AmEx:
\begin{itemize}
    \item Best F1: 0.68827 (MIQUBO, $k{=}150$, D-Wave hybrid solver)
    \item Best Gini: 85.07098\% (MIQUBO, $k{=}150$, D-Wave hybrid solver)
\end{itemize}
 The AmEx working set used here comprises 159 features, versus the 174-feature working set reported in \cite{samanta2025aaai}. This difference stems from feature-level preprocessing: eleven categorical columns (e.g.\ \texttt{B\_30}, \texttt{D\_63}, \texttt{D\_64}) are dropped explicitly rather than label-encoded, since the correlation- and mutual-information-based QUBO methods used throughout this paper are built for continuous numeric relationships, and a 30\%-missingness threshold is applied to the remaining numeric columns.

\section{Methods}
\label{sec:methods}

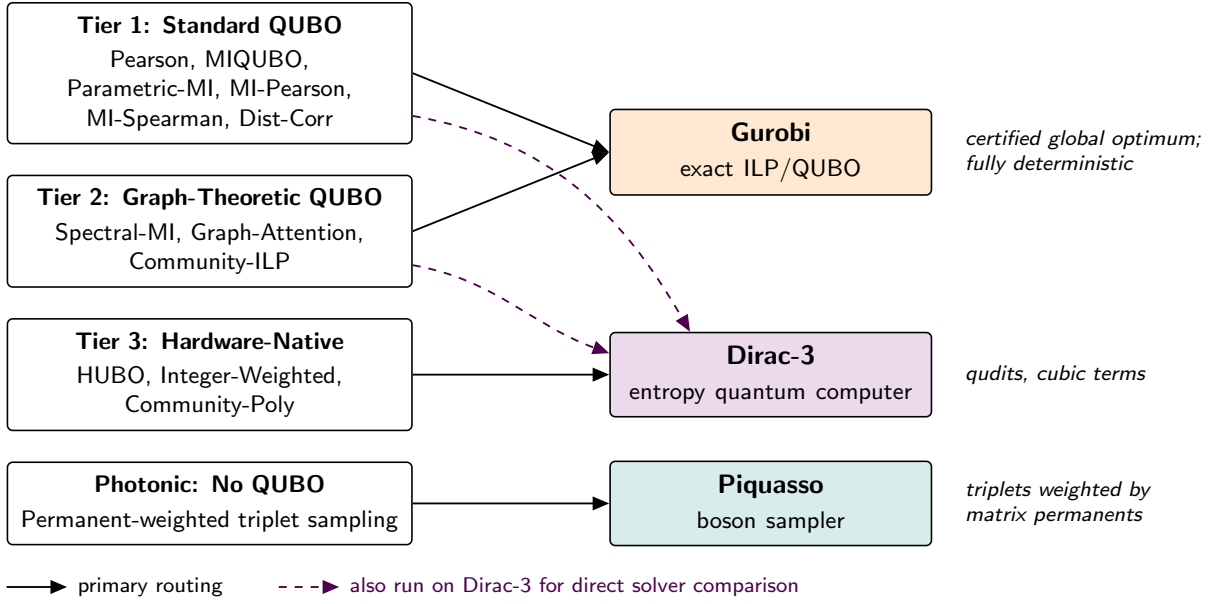
\begin{figure}[!htbp]
\centering
\begin{tikzpicture}[
    font=\sffamily\small,
    box/.style={draw, rounded corners=2pt, align=center, line width=0.6pt, inner sep=5pt,
                execute at begin node={\hyphenpenalty=10000\exhyphenpenalty=10000}},
    method/.style={box, fill=white, text width=5.0cm, font=\sffamily\footnotesize},
    solver/.style={box, text width=3.9cm, font=\sffamily\bfseries\small},
    arrow/.style={-{Triangle[length=2.2mm,width=1.8mm]}, line width=0.7pt},
    alt/.style={arrow, dashed, violet!60!black},
    note/.style={font=\sffamily\scriptsize\itshape, text width=3.3cm, align=left, anchor=west,
                 execute at begin node={\hyphenpenalty=10000\exhyphenpenalty=10000}},
    node distance=4mm
]
\node[method] (tier1) {\textbf{Tier 1: Standard QUBO}\\[2pt]Pearson, MIQUBO, Parametric-MI, MI-Pearson, MI-Spearman, Dist-Corr};
\node[method, below=of tier1] (tier2) {\textbf{Tier 2: Graph-Theoretic QUBO}\\[2pt]Spectral-MI, Graph-Attention, Community-ILP};
\node[method, below=of tier2] (tier3) {\textbf{Tier 3: Hardware-Native}\\[2pt]HUBO, Integer-Weighted, Community-Poly};
\node[method, below=of tier3] (photonic) {\textbf{Photonic: No QUBO}\\[2pt]Permanent-weighted triplet sampling};
\coordinate (sx) at ($(tier1.east)+(2.6,0)$);
\node[solver, fill=orange!18, anchor=west] (gurobi) at ($(sx|-tier1.center)!0.5!(sx|-tier2.center)$) {Gurobi\\[1pt]\footnotesize\mdseries exact ILP/QUBO};
\node[solver, fill=violet!15, anchor=west] (dirac)  at (sx|-tier3.center) {Dirac-3\\[1pt]\footnotesize\mdseries entropy quantum computer};
\node[solver, fill=teal!15, anchor=west]   (photon) at (sx|-photonic.center) {Piquasso\\[1pt]\footnotesize\mdseries boson sampler};
\draw[arrow] (tier1.east) -- (gurobi.west);
\draw[arrow] (tier2.east) -- (gurobi.west);
\draw[arrow] (tier3.east) -- (dirac.west);
\draw[arrow] (photonic.east) -- (photon.west);
\draw[alt] ($(tier1.east)!0.6!(tier1.south east)$) to[out=-10, in=120] ($(dirac.north west)!0.25!(dirac.north east)$);
\draw[alt] ($(tier2.east)!0.6!(tier2.south east)$) to[out=-10, in=160] ($(dirac.west)!0.5!(dirac.north west)$);
\coordinate (nx) at ($(gurobi.east)+(0.3,0)$);
\node[note] at (nx|-gurobi.center) {certified global optimum; fully deterministic};
\node[note] at (nx|-dirac.center) {qudits, cubic terms};
\node[note] at (nx|-photon.center) {triplets weighted by matrix permanents};
\draw[arrow] ($(photonic.south west)+(0,-0.55)$) -- ++(0.8,0) node[right, font=\sffamily\scriptsize] {primary routing};
\draw[alt] ($(photonic.south west)+(3.6,-0.55)$) -- ++(0.8,0) node[right, font=\sffamily\scriptsize] {also run on Dirac-3 for direct solver comparison};
\end{tikzpicture}
\caption{Solver-aware method routing: each tier is assigned to the solver that handles its mathematical structure exactly (solid arrows). Tier 1 and Tier 2 QUBO methods are also run on Dirac-3 for a direct, method-matched comparison (dashed arrows).}
\label{fig:pipeline}
\end{figure}

The central design principle of this benchmark is \emph{solver-aware formulation}. Here, a method refers to a specifc feature-selection formulation together with its associated optimisation objective and each method is assigned to the solver capable of handling its mathematical structure exactly. A \emph{method} is any of the thirteen feature-selection approaches compared: twelve optimization objectives (six Tier 1, three Tier 2 and three Tier 3) and the Piquasso sampler, which has no optimization objective. Figure~\ref{fig:pipeline} summarizes this routing; this section describes the three paradigms and the methods assigned to each.

\subsection{Paradigm 1: Classical Exact Optimization (Gurobi)}

Gurobi \cite{gurobi} solves QUBO and Integer Linear Program (ILP) problems to certified global optimality via branch-and-bound. Its output is deterministic and reproducible, making it the natural reference solver. Two classes of method are assigned here: standard pairwise QUBO (Tier 1) and graph-theoretic QUBO with an optional ILP diversity constraint (Tier 2).

\subsubsection{Tier 1: Standard QUBO Methods}

The general feature-selection QUBO minimizes:
\begin{equation}
    \min_{\mathbf{x} \in \{0,1\}^n} \left[ -\sum_{i=1}^{n} r_i x_i + \alpha \sum_{i<j} \rho_{ij} x_i x_j \right],
    \label{eq:qubo}
\end{equation}
subject to $\sum_i x_i = k$ (enforced as a quadratic penalty). Here $x_i \in \{0,1\}$ indicates selection of feature $i$, $r_i$ is the relevance of feature $i$ to the fraud label, $\rho_{ij}$ is the pairwise redundancy between features $i$ and $j$, and $\alpha > 0$ controls the relevance-redundancy trade-off. Six instantiations differ in how $r_i$ and $\rho_{ij}$ are computed:

\begin{itemize}
    \item \textbf{Pearson}: relevance via $|\rho_{\text{Pearson}}(x_i, y)|$; redundancy via absolute Pearson correlation between features.
    \item \textbf{MIQUBO}: relevance via mutual information $\text{MI}(x_i; y)$; redundancy via pairwise MI between features.
    \item \textbf{Parametric-MI}: relevance via $\text{MI}(x_i; y)$ with parametric density estimation; redundancy via parametric MI.
    \item \textbf{MI-Pearson}: relevance via MI; redundancy via Pearson correlation.
    \item \textbf{MI-Spearman}: relevance via MI; redundancy via Spearman rank correlation. Spearman captures monotone non-linear relationships that Pearson misses, which is relevant for the abstract PCA components V1--V28.
    \item \textbf{Dist-Corr}: relevance and redundancy both measured via distance correlation \cite{szekely2007measuring}, which detects arbitrary statistical dependence, not only linear or monotone relationships.
\end{itemize}

\subsubsection{Tier 2: Graph-Theoretic Methods}

Standard QUBO evaluates features pairwise in isolation. Tier 2 methods model the full feature set as a weighted graph $G = (V, E, W)$, where $V = \{1, \ldots, N\}$ indexes the $N$ candidate features ($N{=}30$ for ULB, $N{=}159$ for AmEx), edges connect all feature pairs, and $W_{ij} = |\rho_{ij}|$ is the absolute pairwise correlation. Global structural properties of $G$ -- unavailable from pairwise inspection alone -- modify the QUBO objective.

\textbf{Spectral-MI} augments the relevance score of each feature by its spectral leverage in $G$. Let $L = D - W$ be the unnormalized graph Laplacian with eigendecomposition $L = U \Lambda U^\top$. The spectral leverage of feature $i$ is:
\begin{equation}
    \ell_i = \sum_{j=1}^{p} \frac{U_{ij}^2}{\lambda_j},
\end{equation}
truncated to the $p$ smallest nonzero eigenvalues. Features with high $\ell_i$ occupy structurally peripheral positions in the correlation graph, indicating low redundancy with the bulk of the feature set. The modified relevance score is:
\begin{equation}
    r_i^{\text{Spectral}} = (1 - \beta)\,\text{MI}(x_i; y) + \beta\,\ell_i,
\end{equation}
with $\beta$ controlling the emphasis on graph structure.

\textbf{Graph-Attention} scales the redundancy penalty between features $i$ and $j$ by their average relevance:
\begin{equation}
    \tilde{\rho}_{ij} = \rho_{ij} \cdot \frac{r_i + r_j}{2},
\end{equation}
so that redundancy between high-relevance features is penalized more heavily than equal redundancy between low-relevance features. This is motivated by attention mechanisms in graph neural networks \cite{velickovic2018graph}.

\textbf{Community-ILP} partitions the feature graph into communities via the Louvain algorithm \cite{blondel2008fast}, identifying clusters of mutually correlated features. A hard diversity constraint -- at least one feature selected from each detected community -- is enforced as an ILP constraint. This guarantees structural diversity in the selected set in a way no soft penalty can, and requires Gurobi's ILP solver. Without this constraint, a QUBO solver could select multiple near-duplicate PCA components from the same cluster and satisfy the cardinality budget while achieving poor coverage.

\subsection{Paradigm 2: Quantum-Entropic Annealing (Dirac-3)}

QCI Dirac-3 \cite{qci2024dirac3} is an entropy-based photonic device that exposes two capabilities unavailable to standard classical solvers: (i) multi-level ``qudit'' variables (integer-valued) and (ii) cubic and higher-order polynomial objective terms. All Tier 1 and Tier 2 QUBO methods are also run on Dirac-3, enabling a direct solver-versus-solver comparison on identical problems. Tier 3 methods are designed to exploit Dirac-3’s native multi-level and higher-order capabilities, avoiding the reformulation overhead that would otherwise be required to express them as binary quadratic models. Each Tier 3 method uses one of these two capabilities: HUBO and Community-Poly keep binary variables but add higher-order terms, while Integer-Weighted keeps a quadratic objective but uses multi-level integer variables.

\subsubsection{Tier 3: Dirac-3-Native Methods}

\textbf{HUBO (Higher-Order Unconstrained Binary Optimization)} extends the standard QUBO (Eq.~\ref{eq:qubo}) with a cubic term penalizing three-way conditional redundancy:
\begin{equation}
    \min_{\mathbf{x}} \left[ -\sum_i r_i x_i + \alpha \sum_{i<j} \rho_{ij} x_i x_j + \gamma \sum_{i<j<\ell} \text{CMI}(i;j|\ell)\, x_i x_j x_\ell \right],
    \label{eq:HUBO}
\end{equation}
where $\text{CMI}(i;j|\ell) = \text{MI}(x_i; x_j \mid x_\ell)$ is the conditional mutual information between features $i$ and $j$ given feature $\ell$ \cite{fleuret2004fast}. This penalizes triplets that are jointly redundant even when each pairwise term is individually acceptable -- a relationship that standard QUBO can only approximate by collapsing three-way terms to pairs.

\textbf{Integer-Weighted Selection} replaces binary feature decisions with graded importance levels $x_i \in \{0, 1, 2, 3\}$ (none, weak, moderate, strong), exploiting Dirac-3's qudit support \cite{qci2024dirac3,nguyen2025entropy}. A total-weight budget $\sum_i x_i \leq B$ replaces the exact cardinality constraint, allowing the solver to allocate importance mass flexibly across features rather than making hard in/out decisions.

\textbf{Community-Poly} replicates the diversity guarantee of Community-ILP on Dirac-3, which has no ILP mode. The hard ``at least one feature per community'' constraint is approximated as a polynomial reward:
\begin{equation}
    + \mu \sum_{c \in \mathcal{C}} \left(1 - \prod_{i \in c}(1 - x_i)\right),
    \label{eq:community_poly}
\end{equation}
where $\mathcal{C}$ is the set of detected feature communities. This polynomial expands to a sum of products, expressible as a higher-order binary objective on Dirac-3.

\subsection{Paradigm 3: Photonic Boson Sampling (Piquasso)}

Piquasso \cite{Piquasso2024processor} requires no QUBO method. The method proceeds as follows. Compute relevance scores $r_i = \text{MI}(x_i; y)$ for all features. Construct a symmetric relevance matrix $A \in \mathbb{R}^{n \times n}$ with $A_{ij} = \sqrt{r_i r_j}$. Pass $A$ as the input transfer matrix of a standard (Fock-state) boson sampling circuit implemented in Piquasso. The output distribution of the circuit assigns probability proportional to $|\text{perm}(A_S)|^2$ to each triplet $S$ of size 3, where $\text{perm}(\cdot)$ is the matrix permanent \cite{aaronson2011computational}. Sample this distribution repeatedly; rank features by selection frequency across samples. The top-$k$ features constitute the selected subset.

The permanent-weighting mechanism rewards features that score well jointly across multiple relevance interactions, not merely on average marginal relevance. This is particularly advantageous at small $k$, where every selected feature must carry maximal independent signal and average-relevance rankings are less discriminative.

\section{Experimental Setup}
\label{sec:experiments}

\subsection{Feature Budget Sweep}
Feature budgets are swept separately for each dataset, reflecting their different dimensionality and published reference configurations.

On ULB, all methods are evaluated at $k \in \{5, 10, 13, 15, 20, 25\}$. This range covers budgets from 17\% to 83\% of the 30 available features, tracing the compression-performance frontier for each method. The selected values are not arbitrary: $k{=}5$ and $k{=}10$ represent aggressive compression with high selection pressure; $k{=}13$ corresponds to the best-Gini configuration reported in \cite{samanta2025aaai}; $k{=}15$ and $k{=}20$ represent moderate compression; and $k{=}25$ is the best-F1 configuration reported there.

On AmEx, all methods are evaluated at $k \in \{25, 50, 75, 100, 125, 150\}$. This range covers budgets from 16\% to 94\% of the 159 available features. The selected values follow the same logic as ULB, scaled to the larger feature space: $k{=}25$ and $k{=}50$ represent aggressive compression; $k{=}75$ and $k{=}100$ represent moderate compression; $k{=}125$ and $k{=}150$ approach the full feature set, with $k{=}150$ corresponding to the best F1 and Gini configuration reported in \cite{samanta2025aaai} on this dataset.

\subsection{Downstream Classifier}
Following \cite{samanta2025aaai}, an XGBoost classifier \cite{chen2016xgboost} is trained on each selected feature subset. On ULB, class-weight balancing adjusts the loss contribution of minority-class samples inversely proportional to class frequency, compensating for the 0.17\% fraud prevalence without requiring oversampling; the same balancing is applied on AmEx to account for its own class imbalance. Performance is evaluated on a held-out test split using F1 and Gini as defined in Section~\ref{sec:dataset}.

\subsection{Classical Pre-Filtering for Dirac-3-Native Methods on AmEx}\label{sec:prefilter}
Tier 3 (Dirac-3-native) methods require cubic objective terms whose
count scales combinatorially with candidate-feature count. At AmEx's full
dimensionality (159 features), the HUBO term count exceeds Dirac-3's native degree-3 problem-size
capacity. To run Tier 3 methods on AmEx, we apply a classical pre-filtering
step: mutual information between each feature and the target is computed
via \texttt{mutual\_info\_classif}, and the top 135 features by MI are
retained (135 is the largest pool size within Dirac-3's degree-3 device
limit). As a sanity check against this cheap top-MI cut, we additionally
compute an mRMR-selected pool of the same size and compare the two pools'
overlap; the top-MI pool is retained only when overlap exceeds 120/135
features (89\%), otherwise the mRMR pool is used instead.

Tier 3's reported AmEx performance is therefore conditional on this
classical pre-selection: unlike Tier 1, Tier 2, and the photonic sampler,
which select from the full 159-feature candidate pool, Tier 3
methods on AmEx select from a 135-feature pool that a classical
MI-based filter has already curated, and results should be read with this
in mind. We report Tier 3 AmEx results at $k \in \{25, 50, 75, 100, 125,
130\}$ (rather than the $\{25,\ldots,150\}$ sweep used elsewhere). The
corresponding all-features result on this 135-feature pool (F1 = 0.8036,
Gini = 91.545) differs slightly from the 159-feature all-features
baseline used elsewhere in this paper (F1 = 0.80217, Gini = 91.57234).

\subsection{Dirac-3 Configuration}

Each Tier 1, Tier 2, and Tier 3 Dirac-3 configuration is evaluated across five independent runs, as described in Section~\ref{sec:dataset}, to characterize run-to-run stochastic variance. Within each run, the solver draws multiple samples per (method, $k$) configuration, and the best-of-sample result for that run is reported: 20 samples per run on ULB, and 5 samples per run on AmEx, reflecting the substantially larger QUBO matrices and longer per-sample solve times at AmEx's higher dimensionality. Bootstrap confidence intervals (95\%, 1000 resamples) over the samples within each run characterize within-run variance, and the spread across the five independent runs (reported as mean $\pm$ 95\% CI throughout Section~\ref{sec:results}) characterizes run-to-run variance, confirming that improvements over Gurobi are statistically meaningful rather than single-sample or single-run outliers.

\subsection{Jaccard Similarity Analysis}

For each (method, $k$, solver-group) combination, pairwise Jaccard similarity between selected feature sets is computed:
\begin{equation}
    J(A, B) = \frac{|A \cap B|}{|A \cup B|},
    \label{eq:jaccard}
\end{equation}
where $J{=}1$ indicates identical selections and $J{=}0$ indicates complete disjointness. Two groups of methods are analyzed separately: (i) the nine Dirac-3 QUBO and Graph methods, and (ii) the Dirac-3 Tier 3 methods (Community-Poly, HUBO, Integer-Weighted). This distinguishes the diversity structure within method classes from diversity across paradigms. Cross-solver similarity between Gurobi and Dirac-3 is computed separately for each directly comparable method.

Because raw Jaccard similarity rises with the relative budget $k/n$ for purely combinatorial reasons, we compare every observed value against the similarity expected between two subsets of size $k$ drawn uniformly at random from $n$ features,
\begin{equation}
    \mathbb{E}[J_{\text{rand}}] = \sum_{i=0}^{k} \frac{\binom{k}{i}\binom{n-k}{k-i}}{\binom{n}{k}} \cdot \frac{i}{2k-i} \;\approx\; \frac{k/n}{2-k/n},
    \label{eq:jaccard_null}
\end{equation}
where the sum runs over the hypergeometric distribution of the overlap size $i$. Agreement above $\mathbb{E}[J_{\text{rand}}]$ indicates that two methods share structure in what they select; agreement at $\mathbb{E}[J_{\text{rand}}]$ indicates overlap no greater than chance. Throughout, 95\% confidence intervals across runs are computed as $\pm 1.96\,s/\sqrt{n_{\text{runs}}}$.

\section{Results}
\label{sec:results}

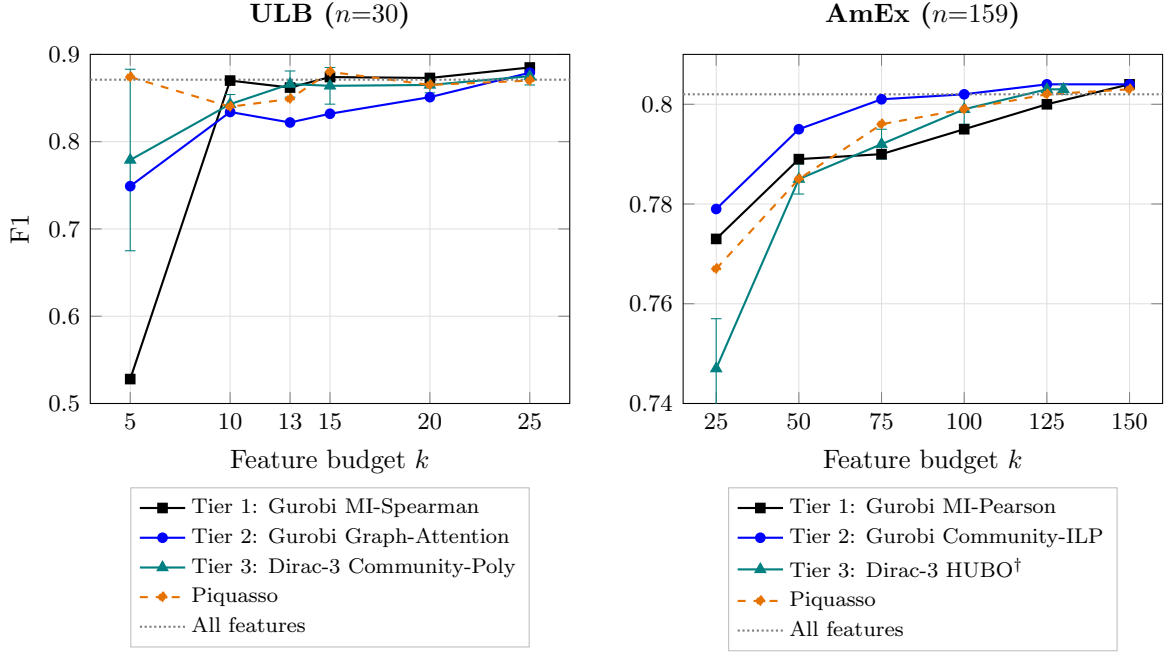
\begin{figure}[htbp]
\centering
\begin{tikzpicture}
\begin{groupplot}[
    group style={group size=2 by 1, horizontal sep=1.5cm},
    width=0.48\linewidth, height=6.2cm,
    xlabel={Feature budget $k$},
    grid=both, grid style={line width=.1pt, draw=gray!25},
    tick label style={font=\footnotesize},
    label style={font=\small},
    title style={font=\small\bfseries},
    error bars/y dir=both, error bars/y explicit,
    legend cell align={left},
    legend style={font=\scriptsize, at={(0.5,-0.24)}, anchor=north, draw=gray!50},
]
\nextgroupplot[title={ULB ($n{=}30$)}, ylabel={F1}, xmin=3, xmax=27, xtick={5,10,13,15,20,25}, ymin=0.50, ymax=0.90]
\addplot[thick, color=black, mark=square*, mark size=1.6pt] coordinates {(5,0.528) (10,0.870) (13,0.862) (15,0.874) (20,0.873) (25,0.885)};
\addlegendentry{Tier 1: Gurobi MI-Spearman}
\addplot[thick, color=blue, mark=*, mark size=1.6pt] coordinates {(5,0.749) (10,0.834) (13,0.822) (15,0.832) (20,0.851) (25,0.879)};
\addlegendentry{Tier 2: Gurobi Graph-Attention}
\addplot[thick, color=teal, mark=triangle*, mark size=2pt] coordinates {(5,0.779) +- (0,0.104) (10,0.843) +- (0,0.011) (13,0.866) +- (0,0.015) (15,0.864) +- (0,0.021) (20,0.865) +- (0,0.009) (25,0.875) +- (0,0.010)};
\addlegendentry{Tier 3: Dirac-3 Community-Poly}
\addplot[thick, dashed, color=orange!90!black, mark=diamond*, mark size=2pt] coordinates {(5,0.874) (10,0.840) (13,0.849) (15,0.880) (20,0.865) (25,0.870)};
\addlegendentry{Piquasso}
\addplot[densely dotted, thick, gray, domain=3:27] {0.871};
\addlegendentry{All features}
\nextgroupplot[title={AmEx ($n{=}159$)}, xmin=15, xmax=160, xtick={25,50,75,100,125,150}, ymin=0.74, ymax=0.81]
\addplot[thick, color=black, mark=square*, mark size=1.6pt] coordinates {(25,0.773) (50,0.789) (75,0.790) (100,0.795) (125,0.800) (150,0.804)};
\addlegendentry{Tier 1: Gurobi MI-Pearson}
\addplot[thick, color=blue, mark=*, mark size=1.6pt] coordinates {(25,0.779) (50,0.795) (75,0.801) (100,0.802) (125,0.804) (150,0.804)};
\addlegendentry{Tier 2: Gurobi Community-ILP}
\addplot[thick, color=teal, mark=triangle*, mark size=2pt] coordinates {(25,0.747) +- (0,0.010) (50,0.785) +- (0,0.003) (75,0.792) +- (0,0.003) (100,0.799) +- (0,0.003) (125,0.803) +- (0,0.001) (130,0.803) +- (0,0.001)};
\addlegendentry{Tier 3: Dirac-3 HUBO$^\dagger$}
\addplot[thick, dashed, color=orange!90!black, mark=diamond*, mark size=2pt] coordinates {(25,0.767) (50,0.785) (75,0.796) (100,0.799) (125,0.802) (150,0.803)};
\addlegendentry{Piquasso}
\addplot[densely dotted, thick, gray, domain=15:160] {0.802};
\addlegendentry{All features}
\end{groupplot}
\end{tikzpicture}
\caption{F1 versus feature budget $k$ for the best solver and method in each tier, together with Piquasso. For each tier, the line shows the solver--method pair with the highest F1 at any budget (Table~\ref{tab:best_results}; ties broken by the smaller $k$, then by the higher mean F1 across budgets), plotted at every budget. Dirac-3 values are five-run means with 95\% CI error bars; Gurobi values are deterministic and Piquasso values were identical across repeated runs. The dotted line is the all-features model. $^\dagger$AmEx Tier 3 uses the 135-feature pre-filtered pool, so its largest budget is $k{=}130$.}
\label{fig:f1_vs_budget}
\end{figure}

Throughout this section, the \emph{best} configuration is the one with the highest F1, the primary metric (Section~\ref{sec:dataset}); Gini is reported alongside it. Ties at three decimals are broken by the smaller $k$, then by the higher mean F1 across all budgets. For Dirac-3, a configuration's F1 is its five-run mean. Table~\ref{tab:best_results} lists, for each solver paradigm, the best configuration over all budgets and the best configuration at the smallest budget. Figure~\ref{fig:f1_vs_budget} follows the best solver and method in each tier, and Piquasso, across all budgets. Section~\ref{sec:results_overview} summarizes the per-method results, with the full per-method discussion in Appendix~\ref{app:detailed}. Section~\ref{sec:cross_solver} then compares Gurobi and Dirac-3 directly on identical methods.

\begin{table}[htbp]
\caption{Best (highest-F1) configuration of each solver paradigm, over all budgets and at the smallest budget ($k{=}5$ on ULB, $k{=}25$ on AmEx). Gurobi entries are deterministic and Piquasso entries were identical across repeated runs; Dirac-3 entries are means $\pm$ 95\% CI over five independent runs. $^\dagger$Tier 3 on AmEx selects from the 135-feature pre-filtered pool (Section~\ref{sec:prefilter}).}
\label{tab:best_results}
\centering
\footnotesize
\setlength{\tabcolsep}{4pt}
\begin{tabular}{lllccc}
\toprule
\textbf{Budget} & \textbf{Paradigm} & \textbf{Method} & \textbf{$k$} & \textbf{F1} & \textbf{Gini (\%)} \\
\midrule
\multicolumn{6}{l}{\textit{ULB}} \\
Best overall & Gurobi (Tier 1)  & MI-Spearman     & 25 & \textbf{0.885} & 96.1 \\
             & Gurobi (Tier 2)  & Graph-Attention & 25 & 0.879 & 95.5 \\
             & Dirac-3 (Tier 1) & MI-Spearman     & 13 & 0.873$\pm$0.023 & 96.1$\pm$0.8 \\
             & Dirac-3 (Tier 2) & Community-ILP   & 15 & 0.870$\pm$0.000 & 95.5$\pm$0.0 \\
             & Dirac-3 (Tier 3) & Community-Poly  & 25 & 0.875$\pm$0.010 & 95.5$\pm$0.6 \\
             & Piquasso         & --              & 15 & 0.880 & 95.4 \\
\cmidrule(lr){1-6}
$k{=}5$      & Gurobi (Tier 1)  & MIQUBO          & 5 & 0.823 & 94.7 \\
             & Gurobi (Tier 2)  & Spectral-MI     & 5 & 0.819 & 93.2 \\
             & Dirac-3 (Tier 1) & MIQUBO          & 5 & 0.804$\pm$0.042 & 94.0$\pm$1.0 \\
             & Dirac-3 (Tier 2) & Community-ILP   & 5 & 0.825$\pm$0.000 & 94.9$\pm$0.0 \\
             & Dirac-3 (Tier 3) & HUBO            & 5 & 0.799$\pm$0.029 & 93.1$\pm$1.8 \\
             & Piquasso         & --              & 5 & \textbf{0.874} & 95.3 \\
\midrule
\multicolumn{6}{l}{\textit{AmEx}} \\
Best overall & Gurobi (Tier 1)  & MI-Pearson      & 150 & 0.804 & 91.5 \\
             & Gurobi (Tier 2)  & Community-ILP   & 125 & 0.804 & 91.6 \\
             & Dirac-3 (Tier 1) & Parametric-MI   & 150 & 0.804$\pm$0.001 & 91.55$\pm$0.03 \\
             & Dirac-3 (Tier 2) & Community-ILP   & 125 & 0.804$\pm$0.000 & 91.56$\pm$0.00 \\
             & Dirac-3 (Tier 3)$^\dagger$ & HUBO  & 125 & 0.803$\pm$0.001 & 91.5$\pm$0.1 \\
             & Piquasso         & --              & 150 & 0.803 & 91.55 \\
\cmidrule(lr){1-6}
$k{=}25$     & Gurobi (Tier 1)  & MIQUBO          & 25 & \textbf{0.789} & 90.5 \\
             & Gurobi (Tier 2)  & Community-ILP   & 25 & 0.779 & 89.6 \\
             & Dirac-3 (Tier 1) & MIQUBO          & 25 & 0.767$\pm$0.013 & 88.45$\pm$1.24 \\
             & Dirac-3 (Tier 2) & Community-ILP   & 25 & 0.783$\pm$0.000 & 89.95$\pm$0.00 \\
             & Dirac-3 (Tier 3)$^\dagger$ & Community-Poly & 25 & 0.754$\pm$0.008 & 87.4$\pm$0.6 \\
             & Piquasso         & --              & 25 & 0.767 & 88.67 \\
\bottomrule
\end{tabular}
\end{table}

\subsection{Overview of Per-Method Results}
\label{sec:results_overview}

Tables~\ref{tab:dirac_qubo} and~\ref{tab:dirac_qubo_amex} give the Dirac-3 F1 of the six Tier 1 methods on each dataset; the corresponding Gurobi values for the budgets compared in Section~\ref{sec:cross_solver} are in Tables~\ref{tab:solver_comparison} and~\ref{tab:solver_comparison_amex}, and the remaining Gurobi and graph-method tables are in Appendix~\ref{app:extra}.

\begin{table}[htbp]
\caption{Dirac-3 -- QUBO Methods, F1 Score by $k$ (mean $\pm$ 95\% CI, $n{=}5$ runs, ULB)}
\label{tab:dirac_qubo}
\centering
\footnotesize
\setlength{\tabcolsep}{3pt}
\begin{tabular}{lccc}
\toprule
\textbf{Method} & \textbf{5} & \textbf{10} & \textbf{13} \\
\midrule
Pearson       & 0.739$\pm$0.038 & 0.791$\pm$0.100 & 0.855$\pm$0.027 \\
MIQUBO        & 0.804$\pm$0.042 & 0.845$\pm$0.019 & 0.856$\pm$0.016 \\
Parametric-MI & 0.488$\pm$0.360 & 0.859$\pm$0.014 & 0.839$\pm$0.022 \\
MI-Pearson    & 0.748$\pm$0.125 & 0.843$\pm$0.025 & 0.845$\pm$0.019 \\
MI-Spearman   & 0.623$\pm$0.192 & 0.842$\pm$0.027 & \textbf{0.873$\pm$0.023} \\
Dist-Corr     & 0.659$\pm$0.155 & 0.825$\pm$0.040 & 0.844$\pm$0.011 \\
\bottomrule
\end{tabular}
\\[4pt]
\begin{tabular}{lccc}
\toprule
\textbf{Method} & \textbf{15} & \textbf{20} & \textbf{25} \\
\midrule
Pearson       & 0.848$\pm$0.013 & 0.863$\pm$0.012 & 0.864$\pm$0.006 \\
MIQUBO        & 0.869$\pm$0.007 & 0.867$\pm$0.011 & 0.870$\pm$0.010 \\
Parametric-MI & 0.855$\pm$0.014 & 0.860$\pm$0.017 & 0.864$\pm$0.017 \\
MI-Pearson    & 0.855$\pm$0.015 & 0.866$\pm$0.008 & 0.869$\pm$0.016 \\
MI-Spearman   & 0.858$\pm$0.009 & 0.865$\pm$0.013 & 0.870$\pm$0.010 \\
Dist-Corr     & 0.856$\pm$0.010 & 0.864$\pm$0.016 & 0.872$\pm$0.009 \\
\bottomrule
\end{tabular}
\end{table}

\begin{table}[htbp]
\caption{Dirac-3 -- QUBO Methods, F1 Score by $k$ (mean $\pm$ 95\% CI, $n{=}5$ runs, AmEx)}
\label{tab:dirac_qubo_amex}
\centering
\footnotesize
\setlength{\tabcolsep}{3pt}
\begin{tabular}{lccc}
\toprule
\textbf{Method} & \textbf{25} & \textbf{50} & \textbf{75} \\
\midrule
Pearson       & 0.755$\pm$0.019 & 0.788$\pm$0.012 & 0.794$\pm$0.006 \\
MIQUBO        & 0.767$\pm$0.013 & 0.789$\pm$0.004 & 0.795$\pm$0.006 \\
Parametric-MI & 0.757$\pm$0.022 & 0.783$\pm$0.008 & 0.792$\pm$0.005 \\
MI-Pearson    & 0.758$\pm$0.016 & 0.785$\pm$0.007 & 0.794$\pm$0.006 \\
Dist-Corr     & 0.746$\pm$0.016 & 0.784$\pm$0.006 & 0.793$\pm$0.006 \\
MI-Spearman   & 0.759$\pm$0.015 & 0.777$\pm$0.010 & 0.794$\pm$0.005 \\
\bottomrule
\end{tabular}
\\[4pt]
\begin{tabular}{lccc}
\toprule
\textbf{Method} & \textbf{100} & \textbf{125} & \textbf{150} \\
\midrule
Pearson       & 0.800$\pm$0.001 & 0.802$\pm$0.002 & 0.803$\pm$0.000 \\
MIQUBO        & 0.799$\pm$0.002 & 0.802$\pm$0.001 & 0.803$\pm$0.000 \\
Parametric-MI & 0.800$\pm$0.003 & 0.801$\pm$0.002 & \textbf{0.804$\pm$0.001} \\
MI-Pearson    & 0.800$\pm$0.004 & 0.802$\pm$0.001 & 0.803$\pm$0.001 \\
Dist-Corr     & 0.797$\pm$0.002 & 0.802$\pm$0.003 & 0.803$\pm$0.000 \\
MI-Spearman   & 0.797$\pm$0.003 & 0.802$\pm$0.001 & 0.803$\pm$0.001 \\
\bottomrule
\end{tabular}
\end{table}

\textbf{ULB.} Method choice matters most at tight budgets. At $k{=}5$, Gurobi's certified optima range from F1 $=0.448$ (Dist-Corr) and $0.459$ (MI-Pearson) to $0.823$ (MIQUBO) (Table~\ref{tab:solver_comparison}), and Dirac-3's five-run means from $0.488$ (Parametric-MI) to $0.804$ (MIQUBO) (Table~\ref{tab:dirac_qubo}). By $k{=}13$ most methods on both solvers lie within a few points of the all-features model (F1 $=0.871$), and Gurobi's MI-Spearman at $k{=}25$ gives the highest deterministic F1 in the benchmark (0.885, Table~\ref{tab:best_results}). On Dirac-3 the same methods show large run-to-run spread at small $k$ (95\% CIs at $k{=}5$ of $\pm0.36$ F1 for Parametric-MI and $\pm0.19$ for MI-Spearman) that narrows below $\pm0.03$ by $k{=}15$; the best Dirac-3 configuration is MI-Spearman at $k{=}13$ (F1 $=0.873\pm0.023$). Community-ILP is fully deterministic on Dirac-3 and is the best Dirac-3 graph method at five of six budgets (Table~\ref{tab:dirac_graph_ulb_f1}).

\textbf{AmEx.} No method on either solver shows a small-$k$ compression advantage. Every method converges to F1 $\approx 0.800$--$0.804$ by $k{=}150$ (Table~\ref{tab:dirac_qubo_amex}), and no configuration exceeds the all-features model (F1 $=0.802$) before $k{=}125$. Community-ILP is the best graph method at every budget on both solvers (Table~\ref{tab:best_results}; Tables~\ref{tab:gurobi_graph_amex} and~\ref{tab:dirac_graph_amex_f1}); its quota over 13 unevenly sized Louvain communities appears to retain informative features from small communities that unconstrained selection omits. Among Tier 1 methods, Gurobi's MIQUBO is best at $k{=}25$ (F1 $=0.789$). Two of Gurobi's certified optima are severely degenerate at $k{=}25$: Dist-Corr (F1 $=0.422$, Gini $=39.6\%$, Table~\ref{tab:solver_comparison_amex}) and Spectral-MI (F1 $=0.644$, Gini $=77.3\%$, Table~\ref{tab:gurobi_graph_amex}), the latter a reversal of its robust small-$k$ behavior on ULB. Dirac-3's run-to-run CIs are much tighter than on ULB, never exceeding $\pm0.022$ F1.

\textbf{Monotonicity.} Within their confidence intervals, the Dirac-3 methods rise with $k$ on both datasets, and every method on both solvers rises with $k$ on AmEx. On ULB, Gurobi's single deterministic results show small reversals (e.g.\ MI-Spearman falls from 0.870 at $k{=}10$ to 0.862 at $k{=}13$; Parametric-MI from 0.743 at $k{=}5$ to 0.701 at $k{=}10$, Table~\ref{tab:gurobi_f1}). The one clearly non-monotonic method is Piquasso on ULB, whose F1 at $k{=}5$ (0.874) is higher than at $k{=}10$, 13, 20 and 25 (Table~\ref{tab:photonic}).

\textbf{Piquasso.} Piquasso is the outlier among the methods. On ULB it is the best method at $k{=}5$ by a wide margin (F1 $=0.874$, against 0.825 for the next-best method, Dirac-3 Community-ILP), slightly above the all-features model with 83\% fewer features, and it gives the best F1 of any method at $k{=}15$ (0.880). On AmEx it has no such advantage (F1 $=0.767$ at $k{=}25$) and tracks the other paradigms at larger budgets. Its rankings were identical across repeated runs, so its results carry no run-to-run CI; Section~\ref{sec:piquasso} discusses the mechanism.

\textbf{Tier 3 (hardware-native).} Table~\ref{tab:tier3} summarizes the three Dirac-3-native methods. On ULB, Community-Poly is the strongest on average (F1 $=0.875\pm0.010$ at $k{=}25$; Gini $=96.2\%$ at $k{=}20$), comparable to Gurobi's best QUBO result, which suggests that the polynomial reward in (Eq.~\ref{eq:community_poly}) approximates the hard Community-ILP constraint closely enough in practice. HUBO has the lowest run-to-run variance of the three (CI half-widths 0.009--0.029) without the top mean at any $k$. Integer-Weighted selection is the least stable Tier 3 method (F1 $=0.779\pm0.109$ at $k{=}10$) with consistently lower Gini (87.3--91.8\%), which we attribute to sensitivity to its total-weight budget $B$. On AmEx, all three climb from F1 $=0.747$--$0.754$ at $k{=}25$ to $0.800$--$0.803$ at $k{=}130$, trailing Gurobi's MIQUBO at $k{=}25$; these results are conditional on the classical pre-filter (Section~\ref{sec:prefilter}).

Appendix~\ref{app:detailed} gives the full discussion of each solver and method family.

\begin{table}[htbp]
\caption{Dirac-3 -- Tier 3 (Hardware-Native) Methods, F1 by $k$ (mean $\pm$ 95\% CI, $n{=}5$ runs). AmEx results use the 135-feature pre-filtered pool (Section~\ref{sec:prefilter}).}
\label{tab:tier3}
\centering
\footnotesize
\setlength{\tabcolsep}{3pt}
\begin{tabular}{lcccccc}
\toprule
\multicolumn{7}{l}{\textit{ULB} ($n{=}30$)} \\
\textbf{Method} & \textbf{5} & \textbf{10} & \textbf{13} & \textbf{15} & \textbf{20} & \textbf{25} \\
\midrule
Community-Poly   & 0.779$\pm$0.104 & 0.843$\pm$0.011 & 0.866$\pm$0.015 & 0.864$\pm$0.021 & 0.865$\pm$0.009 & \textbf{0.875$\pm$0.010} \\
HUBO             & 0.799$\pm$0.029 & 0.843$\pm$0.012 & 0.850$\pm$0.014 & 0.851$\pm$0.020 & 0.857$\pm$0.011 & 0.874$\pm$0.010 \\
Integer-Weighted & 0.720$\pm$0.143 & 0.779$\pm$0.109 & 0.860$\pm$0.017 & 0.845$\pm$0.044 & 0.861$\pm$0.012 & 0.872$\pm$0.005 \\
\midrule
\multicolumn{7}{l}{\textit{AmEx} (135-feature pre-filtered pool)} \\
\textbf{Method} & \textbf{25} & \textbf{50} & \textbf{75} & \textbf{100} & \textbf{125} & \textbf{130} \\
\midrule
Community-Poly   & 0.754$\pm$0.008 & 0.784$\pm$0.004 & 0.796$\pm$0.002 & 0.797$\pm$0.003 & 0.802$\pm$0.001 & 0.803$\pm$0.002 \\
HUBO             & 0.747$\pm$0.010 & 0.785$\pm$0.003 & 0.792$\pm$0.003 & 0.799$\pm$0.003 & 0.803$\pm$0.001 & 0.803$\pm$0.001 \\
Integer-Weighted & 0.748$\pm$0.008 & 0.780$\pm$0.008 & 0.792$\pm$0.004 & 0.795$\pm$0.003 & 0.800$\pm$0.000 & 0.800$\pm$0.000 \\
\bottomrule
\end{tabular}
\end{table}

\subsection{Cross-Solver Comparison}
\label{sec:cross_solver}

Tables~\ref{tab:solver_comparison} and~\ref{tab:solver_comparison_amex} compare Gurobi's deterministic result with the Dirac-3 five-run mean $\pm$ 95\% CI on identical QUBO methods.

Gurobi and Dirac-3 approach the same QUBO differently. Gurobi solves it to certified optimality and returns the same subset on every run, whereas Dirac-3 samples low-energy subsets stochastically, so its output varies between runs and need not coincide with the optimum. Since the QUBO objective only approximates what the classifier needs, the optimal subset is not guaranteed to give the highest F1, and a near-optimal subset from Dirac-3 can sometimes do better. Differences in F1 between the two solvers on the same method therefore say as much about how well the objective reflects classifier performance as about the solvers themselves.

We compare the solvers at two budgets per dataset. On ULB, $k{=}5$ is the smallest budget, where method choice matters most, and $k{=}13$ is the operating point where most methods first approach the all-features model (Section~\ref{sec:discussion}). On AmEx, $k{=}25$ and $k{=}75$ are the budgets closest to these in relative terms (16\% and 47\% of features, against 17\% and 43\% on ULB).

\subsubsection{ULB}

\begin{table}[htbp]
\caption{Gurobi vs.\ Dirac-3 (mean $\pm$ 95\% CI, $n{=}5$) on identical QUBO methods, F1 (ULB). $\Delta$ = Dirac-3 mean $-$ Gurobi. $^{*}$Gurobi's value lies outside the Dirac-3 confidence interval.}
\label{tab:solver_comparison}
\centering
\footnotesize
\setlength{\tabcolsep}{4pt}
\begin{tabular}{lcccc}
\toprule
\textbf{Method} & \textbf{$k$} & \textbf{Gurobi} & \textbf{Dirac-3 (mean$\pm$CI)} & \textbf{$\Delta$} \\
\midrule
Pearson       & 5  & 0.738 & \textbf{0.739$\pm$0.038} & +0.001 \\
MIQUBO        & 5  & \textbf{0.823} & 0.804$\pm$0.042 & $-$0.019 \\
Parametric-MI & 5  & \textbf{0.743} & 0.488$\pm$0.360 & $-$0.255 \\
MI-Pearson    & 5  & 0.459 & \textbf{0.748$\pm$0.125} & +0.289$^{*}$ \\
MI-Spearman & 5  & 0.528 & \textbf{0.623$\pm$0.192} & +0.095 \\
Dist-Corr   & 5  & 0.448 & \textbf{0.659$\pm$0.155} & +0.211$^{*}$ \\
\midrule
Pearson       & 13 & 0.805 & \textbf{0.855$\pm$0.027} & +0.050$^{*}$ \\
MIQUBO        & 13 & 0.836 & \textbf{0.856$\pm$0.016} & +0.020$^{*}$ \\
Parametric-MI & 13 & \textbf{0.851} & 0.839$\pm$0.022 & $-$0.012 \\
MI-Pearson    & 13 & \textbf{0.862} & 0.845$\pm$0.019 & $-$0.017 \\
MI-Spearman & 13 & 0.862 & \textbf{0.873$\pm$0.023} & +0.011 \\
Dist-Corr   & 13 & \textbf{0.848} & 0.844$\pm$0.011 & $-$0.004 \\
\bottomrule
\end{tabular}
\end{table}

On ULB, Dirac-3's mean exceeds Gurobi's result for four of six methods at $k{=}5$ and three of six at $k{=}13$, but most differences are within run-to-run variation. At $k{=}5$, Gurobi's optima for MI-Pearson (F1 $=0.459$) and Dist-Corr (0.448) lie below the entire Dirac-3 interval. At $k{=}13$, Dirac-3 is ahead for Pearson ($+0.050$) and MIQUBO ($+0.020$). Parametric-MI at $k{=}5$ goes the other way ($-0.255$), but its Dirac-3 interval ($\pm0.360$) is too wide for the difference to be meaningful.

The selected features show why certified optimality does not guarantee F1. The four features most correlated with fraud are V10, V12, V14 and V17 (Section~\ref{sec:dependence}). At $k{=}5$, Gurobi's MIQUBO optimum contains all four ($\{$V10, V11, V12, V14, V17$\}$, F1 $=0.823$), while its MI-Pearson optimum ($\{$Time, V2, V3, V25, Amount$\}$, F1 $=0.459$) and MI-Spearman optimum ($\{$Time, V1, V2, V3, Amount$\}$, F1 $=0.528$) contain none of them. Both are certified optimal for their objectives, yet those objectives favor Time, Amount and the leading components V1--V3 over the components most correlated with fraud. Dirac-3's stochastic search does not return these exact optima, and several of its runs land on subsets that contain more of the fraud-relevant components. The best single Dirac-3 run in the benchmark, MI-Spearman at $k{=}13$ (F1 $=0.896$, Table~\ref{tab:best_runs}), selects $\{$Time, V1, V3, V7, V8, V13, V14, V16, V18, V19, V22, V24, Amount$\}$ and exceeds every Gurobi optimum on ULB (the highest is 0.885). This shows Dirac-3 finding a subset that serves the classifier better than the exact optimum found by Gurobi.

\subsubsection{AmEx}

\begin{table}[htbp]
\caption{Gurobi vs.\ Dirac-3 (mean $\pm$ 95\% CI, $n{=}5$) on identical QUBO methods, F1 (AmEx). $k{=}25$ and $k{=}75$ correspond in relative budget to ULB's $k{=}5$ and $k{=}13$ (16\% and 47\% vs.\ 17\% and 43\% of features). $\Delta$ = Dirac-3 mean $-$ Gurobi. $^{*}$Gurobi's value lies outside the Dirac-3 confidence interval.}
\label{tab:solver_comparison_amex}
\centering
\footnotesize
\setlength{\tabcolsep}{4pt}
\begin{tabular}{lcccc}
\toprule
\textbf{Method} & \textbf{$k$} & \textbf{Gurobi} & \textbf{Dirac-3 (mean$\pm$CI)} & \textbf{$\Delta$} \\
\midrule
Pearson       & 25 & \textbf{0.774} & 0.755$\pm$0.019 & $-$0.019 \\
MIQUBO        & 25 & \textbf{0.789} & 0.767$\pm$0.013 & $-$0.022$^{*}$ \\
Parametric-MI & 25 & 0.722 & \textbf{0.757$\pm$0.022} & +0.035$^{*}$ \\
MI-Pearson    & 25 & \textbf{0.773} & 0.758$\pm$0.016 & $-$0.015 \\
Dist-Corr     & 25 & 0.422 & \textbf{0.746$\pm$0.016} & +0.324$^{*}$ \\
MI-Spearman   & 25 & \textbf{0.777} & 0.759$\pm$0.015 & $-$0.018$^{*}$ \\
\midrule
Pearson       & 75 & 0.789 & \textbf{0.794$\pm$0.006} & +0.005 \\
MIQUBO        & 75 & \textbf{0.799} & 0.795$\pm$0.006 & $-$0.004 \\
Parametric-MI & 75 & 0.762 & \textbf{0.792$\pm$0.005} & +0.030$^{*}$ \\
MI-Pearson    & 75 & 0.790 & \textbf{0.794$\pm$0.006} & +0.004 \\
Dist-Corr     & 75 & 0.709 & \textbf{0.793$\pm$0.006} & +0.084$^{*}$ \\
MI-Spearman   & 75 & 0.793 & \textbf{0.794$\pm$0.005} & +0.001 \\
\bottomrule
\end{tabular}
\end{table}

On AmEx the balance depends on the budget. At $k{=}25$, Gurobi's certified solution exceeds the Dirac-3 mean for four of six methods, by 0.015--0.022 F1, whereas at the matched ULB budget Dirac-3 was ahead for four of six. The spread across methods differs sharply between the solvers: Dirac-3's six means at $k{=}25$ span only 0.746--0.767, while Gurobi's certified optima span 0.422--0.789. Dirac-3's results depend little on which method is chosen; Gurobi's depend strongly on it. By $k{=}75$, Dirac-3 is ahead for five of six methods, but for four of them by at most $0.005$ F1. On both datasets, then, most cross-solver differences are small relative to Dirac-3's run-to-run variation, and the clear gaps occur where Gurobi's certified optimum generalizes poorly (Parametric-MI and, above all, Dist-Corr). For Dist-Corr at $k{=}25$ the gap is $+0.324$ F1 (0.746 vs.\ 0.422), the largest in the benchmark and the clearest evidence that certified QUBO optimality does not guarantee downstream performance. The two solvers also share fewer features on AmEx than on ULB relative to chance (Section~\ref{sec:diversity}), yet this extra disagreement does not favor Dirac-3.

\subsection{Best Single-Run Results}
\label{sec:best_single}

\subsubsection{ULB}

Tables~\ref{tab:dirac_qubo}--\ref{tab:solver_comparison_amex} report means over five Dirac-3 runs, but a production deployment ultimately uses a single run's output. Moreover, because Dirac-3 explores multiple near-optimal solutions stochastically, individual runs can produce feature subsets with higher downstream F1 or Gini than the certified optimum returned by Gurobi for the same formulation. Table~\ref{tab:best_runs} therefore reports the best individual run for each of F1 and Gini.

\begin{table}[htbp]
\caption{Best Individual Dirac-3 Run by Metric, vs.\ All-Features Model (ULB)}
\label{tab:best_runs}
\centering
\footnotesize
\setlength{\tabcolsep}{3pt}
\begin{tabular}{lccccc}
\toprule
\textbf{Metric} & \textbf{Method} & \textbf{$k$} & \textbf{Run} & \textbf{Value} & \textbf{$\Delta$ vs.\ All Features} \\
\midrule
Best F1   & MI-Spearman & 13 & 1 & 0.896   & $+0.025$ \\
Best Gini & Dist-Corr   & 25 & 3 & 97.17\% & $+1.61$ pts \\
\bottomrule
\end{tabular}
\\[3pt]
\raggedright\footnotesize All-features model: F1 $= 0.871$; Gini $= 95.56\%$.
\end{table}

The best single-run F1 in the benchmark is Dirac-3 MI-Spearman at $k{=}13$ (F1 $=0.896$), $0.025$ above the all-features model (0.871) with 13 of 30 features. This is also the strongest configuration on average (Table~\ref{tab:dirac_qubo}), so the single-run and mean-based results agree on method and budget, although any individual re-run will vary within the reported CI.

The best single-run Gini was Dist-Corr at $k{=}25$ (Gini $=97.17\%$), $1.61$ points above the all-features model (95.56\%).

\subsubsection{AmEx}

Table~\ref{tab:best_runs_amex} reports the best individual run observed for each of F1 and Gini on AmEx.

\begin{table}[htbp]
\caption{Best Individual Dirac-3 Run by Metric, vs.\ All-Features Model (AmEx)}
\label{tab:best_runs_amex}
\centering
\footnotesize
\setlength{\tabcolsep}{3pt}
\begin{tabular}{lccccc}
\toprule
\textbf{Metric} & \textbf{Method} & \textbf{$k$} & \textbf{Run} & \textbf{Value} & \textbf{$\Delta$ vs.\ All Features} \\
\midrule
Best F1   & Parametric-MI & 150 & 1 & 0.8043  & $+0.002$ \\
Best Gini & MI-Pearson    & 150 & 2 & 91.60\% & $+0.03$ pts \\
\bottomrule
\end{tabular}
\\[3pt]
\raggedright\footnotesize All-features model: F1 $= 0.80217$; Gini $= 91.57234\%$.
\end{table}

The best individual-run F1 (Parametric-MI, $k{=}150$, Run 1, F1 $=0.8043$) and best individual-run Gini (MI-Pearson, $k{=}150$, Run 2, Gini $=91.60\%$) both occur at the largest budget tested, and both sit only marginally above the all-features model (F1 $=0.8022$, Gini $=91.57\%$). This is consistent with the near-monotonic, non-compressive pattern already observed in the mean results (Table~\ref{tab:dirac_qubo_amex}): unlike ULB, where the best single run occurs at a compressed $k{=}13$, AmEx's best individual runs occur at $k{=}150$, the point of least compression tested. No AmEx configuration at small $k$ (25--75) produced a best-run result competitive with its $k{=}150$ counterpart, reinforcing that this dataset does not reward aggressive feature-budget reduction.

\subsection{Piquasso}
\label{sec:piquasso}

Piquasso's sampling is stochastic, but repeated runs returned identical feature rankings at every budget, so unlike Dirac-3 its results show no run-to-run variation.

\subsubsection{ULB}

Table~\ref{tab:photonic} reports full F1 and Gini results for Piquasso across all $k$ values.

\begin{table}[htbp]
\caption{Photonic Boson Sampler -- F1 and Gini (\%) by $k$ (ULB)}
\label{tab:photonic}
\centering
\begin{tabular}{lcccccc}
\toprule
\textbf{Metric} & \textbf{5} & \textbf{10} & \textbf{13} & \textbf{15} & \textbf{20} & \textbf{25} \\
\midrule
F1     & 0.874 & 0.840 & 0.849 & \textbf{0.880} & 0.865 & 0.870 \\
Gini\% & 95.34 & 95.37 & 96.28 & 95.37 & 95.95 & \textbf{96.34} \\
\bottomrule
\end{tabular}
\end{table}

 The all-features baseline for comparison is F1 = 0.871, Gini = 95.56\%.

\textbf{At $k{=}5$, Piquasso outperforms every method on both solvers.} It achieves F1 = 0.874 and Gini = 95.34\%. For reference, the best QUBO result at $k{=}5$ is MIQUBO on Gurobi (F1 = 0.823). Piquasso also slightly exceeds the all-features model (F1 $= 0.874$ vs.\ $0.871$) at $k{=}5$, an 83\% reduction in feature count relative to the full 30-feature set. The advantage at $k{=}5$ is attributable to permanent weighting: the probability assigned to triplet $S$ is proportional to $|\text{perm}(A_S)|^2$, which reflects the joint relevance of all three features simultaneously rather than a sum or product of marginal relevances. At very tight budgets, this joint weighting selects features that are collectively strong, not just individually strong.

\textbf{Performance dips at $k{=}10$ before recovering.} F1 falls from 0.874 at $k{=}5$ to 0.840 at $k{=}10$; Piquasso is the only method whose $k{=}5$ result exceeds its results at $k{=}10$, 13, 20 and 25. The mechanism is structural: the sampler generates triplets ($|S|{=}3$), and the current frequency-counting aggregation step selects features by their marginal frequency across triplet samples. When extending from $k{=}5$ to $k{=}10$, the marginal frequency ranking does not preserve the pairwise complementarity that produced the strong $k{=}5$ result. Features that appear frequently in isolation may be mutually redundant when combined. This is a limitation of the aggregation step, not of the permanent-weighting mechanism itself, and could be addressed by a coverage-maximizing aggregation over sampled triplets.

\textbf{Recovery is complete by $k{=}13$--$15$.} At $k{=}15$, F1 = 0.880 places Piquasso among the top three methods across all paradigms at that budget. At $k{=}25$, Gini = 96.34\% is the highest of any photonic or graph-based method in the benchmark. These results confirm that Piquasso is not merely a small-$k$ heuristic: once the aggregation step encounters sufficient triplet coverage relative to the target budget, performance is competitive with the best QUBO-based methods.

\subsubsection{AmEx}

Table~\ref{tab:photonic_amex} reports full F1 and Gini results for Piquasso on AmEx.

\begin{table}[htbp]
\caption{Photonic Boson Sampler -- F1 and Gini (\%) by $k$ (AmEx)}
\label{tab:photonic_amex}
\centering
\begin{tabular}{lcccccc}
\toprule
\textbf{Metric} & \textbf{25} & \textbf{50} & \textbf{75} & \textbf{100} & \textbf{125} & \textbf{150} \\
\midrule
F1     & 0.767 & 0.785 & 0.796 & 0.799 & 0.802 & \textbf{0.803} \\
Gini\% & 88.67 & 89.97 & 90.92 & 91.23 & 91.40 & \textbf{91.55} \\
\bottomrule
\end{tabular}
\end{table}

 The all-features baseline for comparison is F1 $=0.8022$, Gini $=91.57\%$.

\textbf{Piquasso's small-$k$ advantage on ULB does not transfer to AmEx.} At $k{=}25$ (16\% of features, comparable in relative terms to ULB's $k{=}5$ at 17\%), the sampler reaches only F1 $=0.767$, well below the all-features baseline (0.802) and below Gurobi's best Tier 1 result at the same budget (MIQUBO, F1 $=0.789$).

\textbf{Performance climbs monotonically with $k$ and saturates near the all-features baseline only as $k$ approaches the full feature count.} F1 rises steadily from 0.767 at $k{=}25$ to 0.803 at $k{=}150$ (94\% of available features), matching the baseline only at the largest budget tested; Gini follows the same pattern, reaching 91.55\% at $k{=}150$ against a baseline of 91.57\%. 

\textbf{The permanent-weighting mechanism that gives the sampler its small-$k$ edge on ULB appears not to confer the same advantage on AmEx's larger, denser feature space.} The dependence structure in Section~\ref{sec:dependence} suggests why. On ULB the fraud signal is concentrated in a few components (only V17 and V14 have $|r_y|>0.3$), so a sampler that rewards jointly relevant triplets can find most of it with five features. On AmEx the signal is spread over many features (29 with $|r_y|>0.3$), so a small sampled subset captures only part of it, and performance improves steadily as more features are added.

\section{Subset Diversity: Jaccard Similarity Analysis}
\label{sec:diversity}

Whether distinct solvers or methods explore distinct regions of the feature-subset space, or converge to the same solution, is quantified directly by Jaccard similarity (Eq.~\ref{eq:jaccard}). Where available, we report the mean off-diagonal Jaccard similarity as mean $\pm$ 95\% CI across independent Dirac-3 runs to characterize how stable the diversity pattern is to re-sampling. Every value is read against the random-subset expectation $\mathbb{E}[J_{\text{rand}}]$ of (Eq.~\ref{eq:jaccard_null}), since raw overlap rises with $k/n$ regardless of what the methods select. We report this analysis separately for ULB and AmEx.

All Jaccard values in this section are computed from the feature subsets selected in the runs of Section~\ref{sec:results}: Gurobi's single certified subset for each (method, $k$), and Dirac-3's subset for each (method, $k$, run). Table~\ref{tab:jaccard_extra} collects the supplementary values quoted in the text that are not part of the trend tables.

\begin{table}[htbp]
\caption{Supplementary Jaccard similarities quoted in Section~\ref{sec:diversity}, with random-subset expectation (Eq.~\ref{eq:jaccard_null}). Dirac-3 values are averaged over runs.}
\label{tab:jaccard_extra}
\centering
\footnotesize
\begin{tabular}{llccc}
\toprule
\textbf{Dataset} & \textbf{Comparison} & \textbf{$k$} & \textbf{Jaccard} & \textbf{$\mathbb{E}[J_{\text{rand}}]$} \\
\midrule
ULB  & Gurobi vs.\ Dirac-3, Spectral-MI     & 5  & 0.111 & 0.099 \\
ULB  & Gurobi vs.\ Dirac-3, Graph-Attention & 5  & 0.056 & 0.099 \\
ULB  & Gurobi vs.\ Dirac-3, Dist-Corr       & 5  & 0.118 & 0.099 \\
ULB  & Gurobi vs.\ Dirac-3, Community-ILP   & 13 & 0.529 & 0.282 \\
\midrule
AmEx & Gurobi vs.\ Dirac-3, Pearson         & 25 & 0.060 & 0.087 \\
AmEx & Gurobi vs.\ Dirac-3, MIQUBO          & 25 & 0.153 & 0.087 \\
AmEx & Gurobi Tier 1 methods with each other & 25 & 0.181 & 0.087 \\
AmEx & Gurobi Tier 1 methods with each other & 50 & 0.318 & 0.188 \\
AmEx & Dirac-3 same method across runs & 25 & 0.097 & 0.087 \\
\bottomrule
\end{tabular}
\end{table}

\subsection{Diversity Among QUBO and Graph Methods}

\subsubsection{ULB}

We report the mean off-diagonal Jaccard similarity among the nine Dirac-3 QUBO and Graph methods as a function of $k$, averaged across runs. Full pairwise matrices are run-specific and are not averaged here.

\begin{table}[htbp]
\caption{Mean Off-Diagonal Jaccard Similarity vs.\ $k$ -- QUBO/Graph Methods (mean $\pm$ 95\% CI, ULB), with Random-Subset Expectation}
\label{tab:jaccard_trend_qubograph_ulb}
\centering
\footnotesize
\begin{tabular}{ccc}
\toprule
\textbf{$k$} & \textbf{Mean Jaccard} & \textbf{$\mathbb{E}[J_{\text{rand}}]$} \\
\midrule
5  & 0.096$\pm$0.022 & 0.099 \\
10 & 0.203$\pm$0.006 & 0.207 \\
13 & 0.289$\pm$0.012 & 0.282 \\
15 & 0.333$\pm$0.029 & 0.339 \\
20 & 0.508$\pm$0.021 & 0.503 \\
25 & 0.719$\pm$0.008 & 0.716 \\
\bottomrule
\end{tabular}
\end{table}

Mean pairwise Jaccard similarity rises monotonically with $k$, from $0.096 \pm 0.022$ at $k{=}5$ to $0.719 \pm 0.008$ at $k{=}25$, and the tight CIs at every $k$ indicate this trend is stable across the available re-runs rather than an artifact of any single sampling trajectory. The rise itself is combinatorial: at every tested budget the observed value lies within $0.01$ of the random-subset expectation (Table~\ref{tab:jaccard_trend_qubograph_ulb}). Dirac-3's nine methods therefore overlap with one another no more than randomly drawn subsets of the same size would. We do not report individual pairwise method matrices (e.g., which specific pair is ``most distinct'' at a given $k$) as a multi-run average, since matrix entries are not directly comparable across runs without first establishing a stable method ordering; single-run illustrative matrices for $k{=}5$ and $k{=}25$ are reported in Appendix~\ref{app:jaccard_matrices} instead.

\subsubsection{AmEx}
\label{sec:jaccard_tier12_amex}

We repeat the same analysis for AmEx. Table~\ref{tab:jaccard_trend_qubograph_amex} reports the mean off-diagonal Jaccard similarity as a function of $k$, computed from feature-subset overlap within each of the five independent Dirac-3 runs.

\begin{table}[htbp]
\caption{Mean Off-Diagonal Jaccard Similarity vs.\ $k$ -- QUBO/Graph Methods (mean $\pm$ 95\% CI, AmEx), with Random-Subset Expectation}
\label{tab:jaccard_trend_qubograph_amex}
\centering
\footnotesize
\begin{tabular}{ccc}
\toprule
\textbf{$k$} & \textbf{Mean Jaccard} & \textbf{$\mathbb{E}[J_{\text{rand}}]$} \\
\midrule
25  & 0.097$\pm$0.006 & 0.087 \\
50  & 0.184$\pm$0.002 & 0.188 \\
75  & 0.309$\pm$0.004 & 0.310 \\
100 & 0.462$\pm$0.005 & 0.459 \\
125 & 0.652$\pm$0.005 & 0.648 \\
150 & 0.893$\pm$0.001 & 0.893 \\
\bottomrule
\end{tabular}
\end{table}

\begin{figure}[htbp]
\centering
\begin{tikzpicture}
\begin{axis}[
    width=0.95\linewidth,
    height=6cm,
    xlabel={Relative feature budget $k/n$ (\%)},
    ylabel={Mean off-diagonal Jaccard similarity},
    xmin=0, xmax=100,
    ymin=0, ymax=1,
    grid=both,
    grid style={line width=.1pt, draw=gray!30},
    legend pos=north west,
    legend cell align={left},
    error bars/y dir=both,
    error bars/y explicit,
    tick label style={font=\footnotesize},
    label style={font=\small},
    legend style={font=\footnotesize},
]

\addplot+[
    mark=o, thick, color=blue,
    error bars/y dir=both, error bars/y explicit
]
coordinates {
    (16.7, 0.096) +- (0, 0.022)
    (33.3, 0.203) +- (0, 0.006)
    (43.3, 0.289) +- (0, 0.012)
    (50.0, 0.333) +- (0, 0.029)
    (66.7, 0.508) +- (0, 0.021)
    (83.3, 0.719) +- (0, 0.008)
};
\addlegendentry{ULB (fraud, $n{=}30$)}

\addplot+[
    mark=square*, thick, color=red,
    error bars/y dir=both, error bars/y explicit
]
coordinates {
    (15.7, 0.097) +- (0, 0.006)
    (31.4, 0.184) +- (0, 0.002)
    (47.2, 0.309) +- (0, 0.004)
    (62.9, 0.462) +- (0, 0.005)
    (78.6, 0.652) +- (0, 0.005)
    (94.3, 0.893) +- (0, 0.001)
};
\addlegendentry{AmEx (default, $n{=}159$)}

\addplot[dashed, thick, color=black, domain=1:99, samples=60]
    {(x/100)/(2 - x/100)};
\addlegendentry{Random subsets, $\frac{k/n}{2-k/n}$}

\end{axis}
\end{tikzpicture}
\caption{Mean off-diagonal Jaccard similarity among QUBO/Graph feature-selection
methods as a function of relative feature budget $k/n$, for ULB and AmEx.
Error bars show 95\% CI across five independent Dirac-3 runs (Tables
\ref{tab:jaccard_trend_qubograph_ulb} and
\ref{tab:jaccard_trend_qubograph_amex}). The dashed line is the expected
similarity between two random subsets of the same size (Eq.~\ref{eq:jaccard_null}).
Both datasets lie on this curve at every tested budget, so cross-method
overlap does not separate them.}
\label{fig:jaccard_curve}
\end{figure}
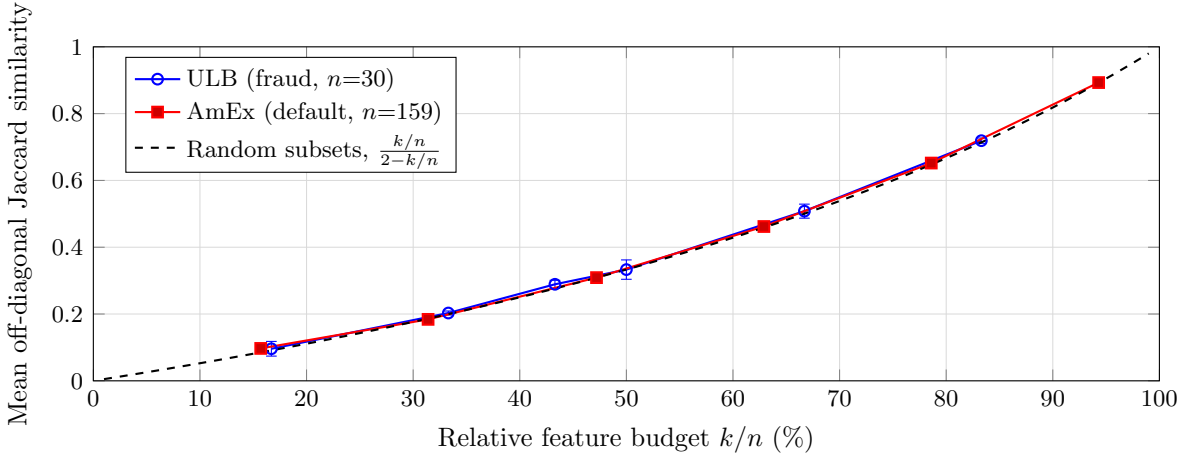

The AmEx trend follows the same monotone, concave-rising shape as ULB, rising from $0.097 \pm 0.006$ at $k{=}25$ to $0.893 \pm 0.001$ at $k{=}150$. Because AmEx's tested budgets (16\%--94\% of 159 features) span almost the identical relative range as ULB's (17\%--83\% of 30 features), the two curves are directly comparable at matched relative budget, and they coincide (Fig.~\ref{fig:jaccard_curve}): AmEx's $k{=}25$ point (16\%, $J{=}0.097$) is indistinguishable from ULB's $k{=}5$ point (17\%, $J{=}0.096$), and AmEx's $k{=}75$ point (47\%, $J{=}0.309$) sits beside ULB's $k{=}13$ point (43\%, $J{=}0.289$). The larger, more correlated structure of AmEx's feature space therefore does \emph{not} lead Dirac-3's methods to converge on a shared core of features, and cross-method overlap on its own does not distinguish the two datasets. What distinguishes them is examined in Section~\ref{sec:overlap_vs_sensitivity}.

The random-level overlap among Dirac-3's methods is not simply a consequence of the methods encoding different objectives. Gurobi's certified solutions to the six Tier 1 methods on AmEx agree with one another well above chance (mean off-diagonal Jaccard $0.181$ at $k{=}25$ versus $\mathbb{E}[J_{\text{rand}}]{=}0.087$; $0.318$ versus $0.188$ at $k{=}50$), and the same Dirac-3 method re-run across its five runs agrees with itself at roughly the random-subset level ($0.097$ at $k{=}25$; Table~\ref{tab:jaccard_extra}). Overlap statistics alone cannot tell us why this happens. One possibility is a near-degenerate objective landscape, in which many subsets have nearly equal QUBO energy; another is that the stochastic solver does not fully converge with five samples per run. Comparing each Dirac-3 solution's QUBO energy against Gurobi's certified optimum would distinguish the two, and we flag this as a limitation (Section~\ref{sec:discussion}).

\subsection{Diversity Among Tier 3 Methods}

\subsubsection{ULB}

\begin{table}[htbp]
\caption{Mean Off-Diagonal Jaccard Similarity vs.\ $k$ -- Tier 3 Methods (mean $\pm$ 95\% CI, $n{=}5$, ULB), with Random-Subset Expectation}
\label{tab:jaccard_trend_tier3}
\centering
\footnotesize
\begin{tabular}{ccc}
\toprule
\textbf{$k$} & \textbf{Mean Jaccard} & \textbf{$\mathbb{E}[J_{\text{rand}}]$} \\
\midrule
5  & 0.116$\pm$0.051 & 0.099 \\
10 & 0.191$\pm$0.024 & 0.207 \\
13 & 0.308$\pm$0.044 & 0.282 \\
15 & 0.351$\pm$0.046 & 0.339 \\
20 & 0.521$\pm$0.048 & 0.503 \\
25 & 0.731$\pm$0.023 & 0.716 \\
\bottomrule
\end{tabular}
\end{table}

Table~\ref{tab:jaccard_trend_tier3} reports the same mean off-diagonal Jaccard trend for the three Tier 3 methods on ULB, across all five runs.

The Tier 3 trend follows the same monotone increase with $k$ as the QUBO/Graph group, rising from $0.116 \pm 0.051$ at $k{=}5$ to $0.731 \pm 0.023$ at $k{=}25$, and, like that group, stays within one confidence interval of the random-subset expectation at every budget. The CIs are wider than for the QUBO/Graph group at every $k$ (e.g., $\pm0.051$ vs.\ $\pm0.022$ at $k{=}5$), which is consistent with there being only three methods in this group rather than nine: with fewer pairs contributing to the mean, a single divergent pair has proportionally more influence on the average, and run-to-run resampling noise shows up more visibly.

\subsubsection{AmEx}

\begin{table}[htbp]
\caption{Mean Off-Diagonal Jaccard Similarity vs.\ $k$ -- Tier 3 Methods (mean $\pm$ 95\% CI, $n{=}5$, AmEx, 135-feature pool), with Random-Subset Expectation for $n{=}135$}
\label{tab:jaccard_trend_tier3_amex}
\centering
\footnotesize
\begin{tabular}{ccc}
\toprule
\textbf{$k$} & \textbf{Mean Jaccard} & \textbf{$\mathbb{E}[J_{\text{rand}}]$} \\
\midrule
25  & 0.105$\pm$0.035 & 0.104 \\
50  & 0.214$\pm$0.024 & 0.229 \\
75  & 0.379$\pm$0.021 & 0.386 \\
100 & 0.589$\pm$0.013 & 0.589 \\
125 & 0.863$\pm$0.006 & 0.862 \\
130 & 0.927$\pm$0.002$^{\dagger}$ & 0.929 \\
\bottomrule
\multicolumn{3}{l}{\scriptsize $^{\dagger}$Four runs; Run 3's Community-Poly output at $k{=}130$ returned}\\
\multicolumn{3}{l}{\scriptsize only 72 features and is excluded.}
\end{tabular}
\end{table}

Table~\ref{tab:jaccard_trend_tier3_amex} reports the mean off-diagonal Jaccard similarity among the three Dirac-3 Tier 3 methods run on AmEx (HUBO, Integer-Weighted, Community-Poly), as mean $\pm$ 95\% CI over five independent runs, on the 135-feature pre-filtered pool described in Section~\ref{sec:prefilter}.

The trend follows the same monotone, concave-rising shape already documented for the AmEx Tier 1/2 group (Section~\ref{sec:jaccard_tier12_amex}) and for ULB's own Tier 3 group above, rising from $0.105 \pm 0.035$ at $k{=}25$ (18.5\% of the 135-feature pool) to $0.927 \pm 0.002$ at $k{=}130$ (96\%), and again tracks the random-subset expectation for a 135-feature pool at every budget. As on ULB, no cross-solver comparison is possible here, since Tier 3 has no Gurobi counterpart. Because the MI pre-filter (Section~\ref{sec:prefilter}) has already reduced the pool to 135 features, the low overlap at small $k$ reflects disagreement over which of those features to keep, not over the full 159-feature space used by Tier 1/2 methods.

\subsection{Cross-Solver Diversity}

\subsubsection{ULB}

Table~\ref{tab:jaccard_cross_solver_ulb} reports the mean Jaccard similarity between Gurobi's selected feature subset and the Dirac-3 selected subset for the same method and $k$, on ULB.

\begin{table}[htbp]
\caption{Mean Cross-Solver Jaccard Similarity (Gurobi vs.\ Dirac-3) vs.\ $k$, Averaged Across Methods and Runs (ULB), with Random-Subset Expectation and Ratio of Observed to Expected}
\label{tab:jaccard_cross_solver_ulb}
\centering
\footnotesize
\begin{tabular}{cccc}
\toprule
\textbf{$k$} & \textbf{Mean Jaccard} & \textbf{$\mathbb{E}[J_{\text{rand}}]$} & \textbf{Ratio} \\
\midrule
5  & 0.195 & 0.099 & 2.0$\times$ \\
10 & 0.301 & 0.207 & 1.5$\times$ \\
13 & 0.396 & 0.282 & 1.4$\times$ \\
15 & 0.475 & 0.339 & 1.4$\times$ \\
20 & 0.616 & 0.503 & 1.2$\times$ \\
25 & 0.798 & 0.716 & 1.1$\times$ \\
\bottomrule
\end{tabular}
\end{table}

Cross-solver Jaccard similarity rises monotonically with $k$, from $0.195$ at $k{=}5$ to $0.798$ at $k{=}25$. It sits clearly above the random-subset expectation at every budget: at $k{=}5$ it is $2.0\times$ the expected $0.099$, and roughly double the within-Dirac-3 QUBO/Graph mean ($0.096$, Table~\ref{tab:jaccard_trend_qubograph_ulb}). Gurobi and Dirac-3 solving the \emph{same} method therefore share more features than Dirac-3's \emph{different} methods share with one another, so on ULB the method, not only the solver, shapes what is selected.

MIQUBO at $k{=}5$ illustrates what this overlap consists of. Gurobi's certified subset is $\{$V10, V11, V12, V14, V17$\}$, which contains all four of the most fraud-correlated components. Dirac-3's runs return, for example, $\{$V4, V7, V12, V14, V20$\}$ (F1 $=0.789$), $\{$V4, V9, V13, V14, V17$\}$ (0.832), $\{$V3, V4, V6, V10, V27$\}$ (0.750) and $\{$V3, V12, V14, V17, V25$\}$ (0.821), with Jaccard 0.11--0.43 against Gurobi's subset. Dirac-3 typically keeps one to three of V10, V12, V14 and V17 and fills the remaining places differently. The run that keeps only V10 scores lowest. F1 is not explained by closeness to the certified optimum: the run with the highest F1 (0.832, above Gurobi's 0.823) shares only two features with it.

The method-level picture is heterogeneous. \textbf{Community-ILP} shows zero run-to-run variance in its cross-solver Jaccard at every $k$ where it is defined (e.g., $0.529$ at $k{=}13$, identical across Dirac-3 runs; Table~\ref{tab:jaccard_extra}): Community-ILP's community detection and integer quota allocation are deterministic given the correlation graph, so Dirac-3 returns the same selected subset on every re-run for this method, unlike the QUBO-based methods, whose run-to-run variance comes from Dirac-3's stochastic annealing. Gurobi's Community-ILP result is unavailable at $k{=}5$: the ILP solver returned an infeasible status under the community-size quotas at this budget, so $k{=}5$ is excluded from the Community-ILP comparison and from the $n_{\text{methods}}{=}8$ average at $k{=}5$ in Table~\ref{tab:jaccard_cross_solver_ulb}.

\textbf{Spectral-MI}, \textbf{Graph-Attention}, and \textbf{Dist-Corr} show the lowest cross-solver similarity at small $k$ (0.111, 0.056, and 0.118 respectively at $k{=}5$; Table~\ref{tab:jaccard_extra}), at or below the random-subset expectation of $0.099$: for these three methods, Gurobi's and Dirac-3's selections at a tight budget are no more alike than two random subsets. This is consistent with the elevated within-Dirac-3 variance already noted for Dist-Corr at small $k$ in Section~\ref{sec:results}: when a method's QUBO landscape is itself unstable to re-sampling, it is unsurprising that two different solvers, sampling that landscape differently, diverge even further from one another.

\subsubsection{AmEx}

Table~\ref{tab:jaccard_cross_solver_amex} reports the mean cross-solver Jaccard similarity on AmEx, computed across the six directly compared Tier 1 methods (Pearson, MIQUBO, Parametric-MI, MI-Pearson, Dist-Corr, MI-Spearman; the same six as Table~\ref{tab:solver_comparison_amex}) and all five Dirac-3 runs.

\begin{table}[htbp]
\caption{Mean Cross-Solver Jaccard Similarity (Gurobi vs.\ Dirac-3) vs.\ $k$, Six Tier 1 Methods (mean $\pm$ 95\% CI over five Dirac-3 runs, AmEx), with Random-Subset Expectation and Ratio of Observed to Expected}
\label{tab:jaccard_cross_solver_amex}
\centering
\footnotesize
\begin{tabular}{cccc}
\toprule
\textbf{$k$} & \textbf{Mean Jaccard} & \textbf{$\mathbb{E}[J_{\text{rand}}]$} & \textbf{Ratio} \\
\midrule
25  & 0.112$\pm$0.007 & 0.087 & 1.3$\times$ \\
50  & 0.217$\pm$0.019 & 0.188 & 1.2$\times$ \\
75  & 0.336$\pm$0.011 & 0.310 & 1.1$\times$ \\
100 & 0.470$\pm$0.010 & 0.459 & 1.0$\times$ \\
125 & 0.663$\pm$0.009 & 0.648 & 1.0$\times$ \\
150 & 0.900$\pm$0.002 & 0.893 & 1.0$\times$ \\
\bottomrule
\end{tabular}
\end{table}

As with the within-Dirac-3 trend, cross-solver Jaccard on AmEx rises monotonically with $k$, from $0.112$ at $k{=}25$ to $0.900$ at $k{=}150$. It stays only slightly above the random-subset expectation throughout ($1.3\times$ at $k{=}25$, falling to near parity by $k{=}100$). Two individual methods illustrate the range. For \textbf{Pearson} at $k{=}25$, Gurobi and Dirac-3 share only 1--4 features across the five runs (mean $J{=}0.060$, below the random expectation; Table~\ref{tab:jaccard_extra}), while by $k{=}150$ the two solutions share 141--143 of 150 selected features ($J{=}0.887$--$0.911$). \textbf{MIQUBO} shows the highest cross-solver agreement of the six methods at $k{=}25$ (mean $J{=}0.153$, $1.8\times$ the random expectation).

Compared at matched relative budget, the two datasets differ in the direction opposite to what their raw feature redundancy might suggest: cross-solver agreement is \emph{lower} relative to chance on AmEx ($1.3\times$ at 16\% of features) than on ULB ($2.0\times$ at 17\%). On both datasets, cross-solver similarity exceeds within-Dirac-3 cross-method similarity at small $k$ ($0.112$ vs.\ $0.097$ on AmEx; $0.195$ vs.\ $0.096$ on ULB), but on AmEx the margin is small. More disagreement between the solvers, however, does not translate into a Dirac-3 advantage on AmEx: as Table~\ref{tab:solver_comparison_amex} shows, Gurobi's exact solution wins at four of six methods at $k{=}25$. How much the two solvers disagree does not, by itself, predict which one comes out ahead.

\subsection{Overlap Versus Performance Sensitivity}
\label{sec:overlap_vs_sensitivity}

If cross-method overlap is the same on both datasets and cross-solver overlap does not predict the winner, the contrast in solver sensitivity between ULB and AmEx must come from elsewhere. The relevant quantity is not how much subsets overlap but how much it \emph{costs} to pick a different subset. Table~\ref{tab:f1_spread} compares the spread of Dirac-3's mean F1 across the nine Tier 1/2 methods at matched relative budgets.

\begin{table}[htbp]
\caption{Spread of Dirac-3 Mean F1 Across Nine Tier 1/2 Methods at Matched Relative Budget}
\label{tab:f1_spread}
\centering
\footnotesize
\setlength{\tabcolsep}{3pt}
\begin{tabular}{lcccc}
\toprule
 & \multicolumn{2}{c}{\textbf{ULB}} & \multicolumn{2}{c}{\textbf{AmEx}} \\
\cmidrule(lr){2-3}\cmidrule(lr){4-5}
\textbf{$k/n$ ($k$ ULB / AmEx)} & \textbf{Range} & \textbf{s.d.} & \textbf{Range} & \textbf{s.d.} \\
\midrule
$\approx$16\% (5 / 25)   & 0.488--0.825 & 0.110 & 0.746--0.783 & 0.011 \\
$\approx$45\% (13 / 75)  & 0.839--0.873 & 0.010 & 0.792--0.795 & 0.001 \\
$\approx$80\% (25 / 125) & 0.863--0.872 & 0.003 & 0.801--0.804 & 0.001 \\
\bottomrule
\end{tabular}
\end{table}

At both $\approx$16\% and $\approx$45\% of features, the spread across methods is about ten times larger on ULB than on AmEx. Run-to-run spread tells the same story: Dirac-3's 95\% confidence intervals on ULB at $k{=}5$ reach $\pm0.19$ F1 (MI-Spearman), $\pm0.25$ (Spectral-MI) and $\pm0.36$ (Parametric-MI), but never exceed $\pm0.02$ on AmEx at any budget. On AmEx, the 45 Dirac-3 subsets at $k{=}25$ (nine methods, five runs) barely overlap, yet all reach F1 between 0.725 and 0.783. The dependence structure in Section~\ref{sec:dependence} accounts for this. AmEx's default signal is spread over many features (29 with $|r_y|>0.3$), so most subsets contain enough of it and are nearly interchangeable. ULB's fraud signal is concentrated in a few uncorrelated components (two with $|r_y|>0.3$), which a subset either captures or misses. This is what gives solver choice room to matter on ULB and little room on AmEx.

This spread describes the typical subset, not every method. Certified optima can still fail badly on either dataset: Gurobi's Tier 1/2 F1 spans 0.448--0.823 on ULB at $k{=}5$ and 0.422--0.789 on AmEx at $k{=}25$, driven by individual methods (Dist-Corr and MI-Pearson on ULB; Dist-Corr and Spectral-MI on AmEx). Because the spread in Table~\ref{tab:f1_spread} could be estimated from a handful of inexpensive subsets before any specialized solver is run, it is a natural candidate for a pre-benchmark indicator of when solver choice is likely to matter; validating it against F1 over random subsets on further datasets is left to future work.

\section{Discussion}
\label{sec:discussion}

\subsection{The Solver-Method Alignment Principle}

Our results suggest that the choice of solver and the choice of method should not be made independently. Gurobi's certified optimality is valuable when the QUBO or ILP objective is a faithful proxy for downstream performance, since the solution is then both optimal and reproducible. In all Tier 1 and Tier 2 methods, however, the objective is only an approximation: Gurobi still finds its optimum, but that optimum need not be the best subset for the classifier. Dirac-3's stochastic search can land on subsets that are QUBO-suboptimal but perform better, which is the most likely reading of the Pearson $k{=}13$ result on ULB.

AmEx shows how far this gap can go. On ULB, the worst case still leaves a usable model: Gurobi's Dist-Corr solution at $k{=}5$ reaches only F1 $=0.448$, but its Gini of 81.8\% shows that it ranks transactions far better than chance. On AmEx, the same method at a comparable budget ($k{=}25$) gives F1 $=0.422$ and Gini $=39.6\%$, much closer to a random ranking. Together with Dist-Corr at $k{=}50$ and Spectral-MI at $k{=}25$, it is one of only three AmEx results in the benchmark with F1 below $0.70$, all of them certified Gurobi optima. A certified optimum can therefore perform poorly while appearing rigorously solved, so exact-solver certification should not be treated as a guarantee of downstream quality.

For Tier 3 methods, Dirac-3 is the only option rather than a heuristic alternative. On ULB it applies directly. On AmEx, the HUBO term count exceeds Dirac-3's native degree-3 capacity, so Tier 3 runs on a 135-feature pool pre-selected by a classical MI filter (Section~\ref{sec:prefilter}). This recovers usable results, but part of the selection is then made by a classical heuristic rather than by the hardware-native method. Hardware capacity must therefore be checked against problem size, and any pre-filtering needed to fit the device should be reported as part of the method.

\subsection{F1 and Gini Measure Different Properties}

\subsubsection{ULB}

On ULB, the configurations that maximize F1 and Gini are generally different. Gurobi's best F1 (MI-Spearman, $k{=}25$, F1 $=0.885$) comes with a Gini of 96.1\%, below the 97.0\% the same method reaches at $k{=}20$, and the best individual Dirac-3 runs for F1 (MI-Spearman, $k{=}13$) and for Gini (Dist-Corr, $k{=}25$) differ in both method and budget (Table~\ref{tab:best_runs}). A subset can rank transactions well by fraud risk without being the best choice for precision-recall balance at a fixed decision threshold, and vice versa. In production, the relevant metric depends on how the model output is used: risk-ranking models (e.g., for analyst review queues) should optimize Gini; threshold-based alert systems should optimize F1. The choice should be made before selecting a feature-selection method, because the same method at the same $k$ can rank differently on the two metrics.

\subsubsection{AmEx}

On AmEx, F1 and Gini largely agree: the best single-run F1 (Dirac-3 Parametric-MI, $k{=}150$) and the best single-run Gini (Dirac-3 MI-Pearson, $k{=}150$) occur at the same budget and differ by less than 0.001 F1 and about 0.05 Gini points. The choice between the two metrics therefore matters much less on AmEx than on ULB.

\subsection{The Small-$k$ Operating Point}

\subsubsection{ULB: The $k{=}13$ Sweet Spot}

Across Tier 1, Tier 2, and Tier 3 methods, the $k{=}13$ configuration consistently represents the best compression-performance trade-off on ULB. The improvement in F1 from $k{=}10$ to $k{=}13$ is large (typically 0.02--0.06 F1 points); the improvement from $k{=}13$ to $k{=}25$ is small (typically 0.002--0.015 F1 points). Reducing from 30 features to 13 provides a 57\% reduction in input dimensionality -- with all the associated benefits in scoring speed, data pipeline complexity, and model governance -- at a cost of less than 0.01 F1 on average relative to using 25 features.

\subsubsection{AmEx: No Comparable Sweet Spot}

AmEx has no analogous operating point. On AmEx, compression never improves meaningfully on the full model: the only configurations that exceed the all-features F1 of $0.802$ do so at $k{=}125$--$150$ (78--94\% of features), and by at most $0.002$. F1 and Gini otherwise rise with $k$ across every paradigm (Section~\ref{sec:results}, Fig.~\ref{fig:f1_vs_budget}). Gains do diminish, with most methods within about $0.01$ F1 of their best by $k{=}75$, but the curve flattens gradually rather than peaking at a small budget.

A practitioner who applied ULB's result as a general rule would therefore lose performance on AmEx. At the budget matching ULB's $k{=}13$ in relative terms ($k{=}75$, 47\% of features), AmEx methods remain 0.005--0.01 F1 below the full model, and at $k{=}25$ (16\%) the shortfall is 0.015--0.06 F1. The reason is structural rather than solver-specific (Section~\ref{sec:dependence}). AmEx's default signal is spread across dozens of moderately informative features, so aggressive compression discards real signal; ULB's fraud signal is concentrated in a few uncorrelated principal components, so a small subset that contains them loses little.

Piquasso shows the same contrast most sharply. On ULB it reaches its second-best F1 at the smallest budget ($k{=}5$, 0.874), the best of any method there, because sampling jointly relevant triplets is an efficient way to find a few concentrated signal carriers. On AmEx the same sampler gives F1 $=0.767$ at $k{=}25$ and improves steadily with $k$, like every other method. Whether Piquasso's ULB advantage survives repeated sampling runs remains to be tested.

\subsection{Limitations}

The ULB dataset covers only 48 hours of European card transactions from 2013. Temporal generalization -- to different periods, geographies, or fraud patterns -- is not assessed here. 

Second, Tier 3 (Dirac-3-native) methods on AmEx run on a 135-feature pool pre-selected by a classical MI filter rather than on the full 159 features, because HUBO's term count at full dimensionality exceeds Dirac-3's native capacity (Section~\ref{sec:prefilter}). The hardware-native comparison between the two datasets is therefore not on identical footing, and AmEx Tier 3 results should be read as conditional on the pre-filter. Third, Dirac-3's cross-method Jaccard similarity lies at the random-subset expectation on both datasets, and the same method re-run on AmEx agrees with itself only at that level (Section~\ref{sec:diversity}). We have not compared each Dirac-3 solution's QUBO energy against Gurobi's certified optimum, so we cannot yet distinguish near-degenerate objective landscapes from incomplete convergence of the stochastic solver. Fourth, our AmEx working set (159 features) differs from the 174-feature set used in \cite{samanta2025aaai}, owing to stricter categorical-column and missingness-threshold filtering, so the reference values reported in that study are not directly comparable with ours; for this reason we cite them for context only and do not benchmark against them.

\section{Conclusion}
\label{sec:conclusion}

We benchmarked an exact classical solver (Gurobi), a photonic entropy computer (QCI Dirac-3), and a photonic boson sampler (Piquasso) on thirteen feature-selection methods across two financial datasets of different scale and structure, and tested whether conclusions drawn on one dataset carry over to the other.

On ULB, moderate compression pays off. Dirac-3 MI-Spearman at $k{=}13$ reaches a mean F1 of $0.873 \pm 0.023$ (best run $0.896$), matching the all-features model with 57\% fewer features. Piquasso is the outlier: with only five features it slightly exceeds the all-features model and outperforms every other method at that budget by about 0.05 F1, an advantage it does not show on AmEx. On AmEx, it does not: F1 rises with $k$ for every paradigm and only matches the full model at $k{=}125$--$150$. The compression result from ULB therefore does not transfer.

Solver choice matters less than the method. Most differences between Gurobi and Dirac-3 on the same method are within Dirac-3's run-to-run variation. The clear gaps occur where Gurobi's certified optimum generalizes poorly, most sharply for Dist-Corr on AmEx, where the exact solution reaches F1 $=0.422$ and Gini $=39.6\%$ against a Dirac-3 mean of $0.746$. Certified QUBO optimality is not, on its own, a guarantee of downstream performance.

The distinction between the two datasets is in how much the choice of subset matters. At matched budget, F1 varies about ten times more across methods on ULB than on AmEx, even though the selected subsets overlap at close to the random-subset level on both. The data explain why: ULB's fraud signal is concentrated in a few uncorrelated components, while AmEx's default signal is spread across many features. This suggests that solver and method choices matter most on compact feature spaces with concentrated signal, and that this sensitivity could be estimated from inexpensive subsets before running a full benchmark.

Several questions remain open. We have not compared Dirac-3's solutions against Gurobi's certified QUBO energies, which is needed to tell near-degenerate objectives apart from incomplete convergence. Tier 3 on AmEx relies on a classical pre-filter to fit Dirac-3's capacity. Testing F1 spread over random subsets as a pre-benchmark indicator, and extending the comparison to further datasets, would show whether the pattern observed here generalizes.

\section*{Acknowledgment}
This research was supported by the Singapore Ministry of Education (MOE) Academic Research Fund (AcRF) Tier 1 grant (Proposal ID:24-SIS-SMU-038). The work was supported by Quantum Computing Inc. through their in-kind contributions. We would also like to thank QuiX for their support. SMU Microsoft
Copilot was used throughout this document to support grammar and language clarity.
The authors have no competing interests to declare that are
relevant to the content of this article.


\bibliographystyle{unsrt}

\bibliography{references}

\clearpage
\appendix
\renewcommand{\topfraction}{0.9}
\renewcommand{\bottomfraction}{0.8}
\renewcommand{\textfraction}{0.05}
\renewcommand{\floatpagefraction}{0.75}
\setcounter{topnumber}{4}
\setcounter{bottomnumber}{3}
\setcounter{totalnumber}{6}

\section*{Appendices}

The appendices provide supporting material for Sections~\ref{sec:results} and~\ref{sec:diversity}. Appendix~\ref{app:detailed} gives the full per-method discussion summarized in Section~\ref{sec:results_overview}. Appendix~\ref{app:extra} collects the per-method F1 and Gini tables that are not shown in the main text. Appendix~\ref{app:jaccard_matrices} reports single-run pairwise Jaccard matrices that illustrate the multi-run averages in Section~\ref{sec:diversity}.

\section{Detailed Per-Method Results}
\label{app:detailed}

This appendix gives the full per-method discussion summarized in Section~\ref{sec:results_overview}. The underlying F1 and Gini tables are collected in Appendix~\ref{app:extra}.

\subsection{Gurobi: Standard QUBO Methods}

\paragraph{ULB.}

Table~\ref{tab:gurobi_f1} reports F1 across the six Tier 1 methods on Gurobi; the corresponding Gini values are in Table~\ref{tab:gurobi_gini}.

\textbf{MI-Spearman is the strongest Tier 1 method on both metrics.} At $k{=}25$ it achieves F1 = 0.885, the highest F1 of any configuration in the benchmark and 0.014 above the all-features model (0.871). At $k{=}20$ it achieves Gini = 97.0\%, the highest Gini of any Tier 1 method. Spearman-based redundancy captures monotone non-linear relationships between the PCA components, which Pearson misses by construction.

\textbf{Several methods degrade sharply at $k{=}5$.} MI-Pearson falls to F1 = 0.459 and Dist-Corr to 0.448, below the all-features baseline of 0.871. Both methods use redundancy metrics that become unreliable at very tight budgets: Pearson misses non-linear relationships that matter when only five features are available, while distance correlation is sensitive to noise in the redundancy matrix when the feature count is small. MIQUBO is the most stable at $k{=}5$ (F1 = 0.823), consistent with mutual information being a more complete dependence measure.

\textbf{Parametric-MI recovers quickly after $k{=}10$.} It achieves Gini = 96.4\% at $k{=}13$ and 96.8\% at $k{=}25$, the best Gini among Tier 1 methods across most of the $k$ range. Its F1 at $k{=}10$ (0.701) is the lowest of any method at that budget, suggesting that its parametric density estimator requires a minimum of approximately 13 features to produce reliable scores.

\paragraph{AmEx.}

Table~\ref{tab:gurobi_f1_amex} reports F1 across the same six Tier 1 methods on Gurobi for AmEx; the corresponding Gini values are in Table~\ref{tab:gurobi_gini_amex}. The all-features baseline on this dataset is F1 $=0.80217$, Gini $=91.57234\%$.

\textbf{MIQUBO is the strongest Tier 1 method at small $k$.} At $k{=}25$ it achieves F1 = 0.789, the highest of any method at that budget by a margin of at least 0.012 over every other Tier 1 method, and only 0.013 below the all-features baseline (0.802) despite using just 16\% of candidate features. Unlike ULB, no method exceeds the all-features baseline at any $k$ tested; the closest approaches are Pearson and MI-Pearson at $k{=}150$ (F1 $=0.804$, marginally above baseline).

\textbf{Distance correlation's exact QUBO optimum is severely degenerate at small $k$.} At $k{=}25$, Dist-Corr collapses to F1 $=0.422$ and Gini $=39.6\%$, far worse than any Tier 1 result on ULB at any budget (ULB's worst case, MI-Pearson at $k{=}5$, still reaches F1 $=0.459$ with a Gini well above chance). Gini near 40\% indicates the certified QUBO optimum for this method is producing a near-random risk ranking on this dataset at this budget; this is the clearest instance in the benchmark of a certified QUBO optimum actively harming downstream classifier performance rather than merely being suboptimal, and is discussed further in Section~\ref{sec:discussion}.

\textbf{Convergence to baseline is gradual and near-monotonic for every method.} All six Tier 1 methods climb steadily with $k$ and land within 0.002--0.004 F1 of one another by $k{=}150$ (0.801--0.804), with no method showing the sharp small-$k$ compression advantage seen on ULB. Parametric-MI and Dist-Corr, the two weakest methods at $k{=}25$, both recover fully by $k{=}150$, suggesting their redundancy metrics require a larger feature pool on this dataset to stabilize than the corresponding ULB budgets required.

\subsection{Gurobi: Graph-Theoretic Methods}

\paragraph{ULB.}

Table~\ref{tab:gurobi_graph} reports F1 and Gini for the three Tier 2 methods on Gurobi.

\textbf{Community-ILP achieves the best Gini of any graph method.} Its maximum of 96.9\% at $k{=}15$ reflects the hard diversity constraint: by guaranteeing at least one feature per Louvain community, it prevents the solver from concentrating selections within a single cluster of correlated PCA components. This improves risk-ranking quality (Gini) more than precision-recall balance (F1), which is consistent with diversity benefiting the full score distribution rather than the threshold-specific cut.

\textbf{Spectral-MI is robust at small $k$.} At $k{=}5$ it reaches F1 = 0.819, the best of any graph method at that budget, and avoids the sharp degradation seen in MI-Pearson and Dist-Corr. Structurally peripheral features, those with high spectral leverage, tend to carry information that the majority of the feature set does not, making them individually strong choices at tight budgets.

\textbf{Graph-Attention favors large $k$.} Its best F1 (0.879 at $k{=}25$) and best Gini (96.5\% at $k{=}20$) both occur at moderate-to-large budgets. This is consistent with the method's mechanism: relevance-weighted redundancy penalties require enough features in the selection for the attention weighting to differentiate meaningfully between high- and low-relevance pairings.

\paragraph{AmEx.}

Table~\ref{tab:gurobi_graph_amex} reports the corresponding F1 and Gini results on AmEx.

\textbf{Spectral-MI collapses sharply at $k{=}25$ on AmEx, unlike on ULB.} Its exact Gurobi solution reaches only F1 $=0.644$, Gini $=77.3\%$, the weakest result of any Gurobi method (Tier 1 or Tier 2) at any budget on either dataset. This is a striking reversal of the ULB pattern, where Spectral-MI was the most robust graph method at small $k$ (F1 $=0.819$ at $k{=}5$, Section~\ref{sec:results}). On AmEx's larger, denser correlation graph, spectral leverage in the Laplacian appears to select structurally peripheral features that carry comparatively little classifier-relevant signal at tight budgets; performance recovers steadily and reaches F1 $=0.800$ by $k{=}150$.

\textbf{Community-ILP is the strongest method (Tier 1 or Tier 2) at every budget on AmEx.} At $k{=}25$ it reaches F1 $=0.779$, Gini $=89.6\%$, trailing only Gurobi's best Tier 1 result (MIQUBO, F1 $=0.789$) by one point while already exceeding every other graph or Tier 1 method at that budget. By $k{=}75$ it matches the all-features baseline (F1 $=0.801$ vs. $0.802$), and it is the only graph method to do so before $k{=}125$. The Louvain community structure on AmEx's 159 features detects 13 communities of markedly uneven size (ranging from 2 to 42 features), and the resulting quota constraint appears to force inclusion of informative features from small, otherwise-underrepresented communities that unconstrained QUBO or spectral selection would omit.

\textbf{Graph-Attention sits between the two}, tracking close to Community-ILP at small $k$ (F1 $=0.743$ at $k{=}25$) and modestly exceeding the all-features baseline by $k{=}150$ (F1 $=0.801$). Its relevance-weighted redundancy penalty behaves consistently with its ULB counterpart: performance improves steadily rather than peaking at any particular budget.

\subsection{Dirac-3: QUBO Methods}

\paragraph{ULB.}

Table~\ref{tab:dirac_qubo} and Table~\ref{tab:dirac_qubo_gini} report F1 and Gini respectively for Dirac-3 QUBO methods across all $k$ values, as the mean $\pm$ 95\% confidence interval (CI) over five independent Dirac-3 runs. Each run draws its own best-of-20-sample solution per (method, $k$) pair; the spread across runs reflects re-run variance in the stochastic annealing process, not measurement noise.

\textbf{On average, Dirac-3 MI-Spearman at $k{=}13$ is the strongest QUBO configuration in the benchmark}, with mean F1 $= 0.873 \pm 0.023$ across five runs, an average gain of $+0.011$ over Gurobi's exact solution at the same $k$ (F1 $= 0.862$; see Table~\ref{tab:solver_comparison}). The width of the CI matters here: it indicates run-to-run variance of roughly $\pm$0.02 F1 at this $k$, so any single re-run could plausibly land below Gurobi's result even though the average favors Dirac-3. On Gini, MI-Spearman peaks at a mean of $96.1\% \pm 0.8$ at $k{=}13$ (Table~\ref{tab:dirac_qubo_gini}).

\textbf{The gap between Dirac-3 and Gurobi narrows at large $k$.} At $k{=}25$, mean MI-Spearman F1 on Dirac-3 ($0.870\pm0.010$) is statistically indistinguishable from Gurobi's exact result (0.885); the CI does not exclude Gurobi's value lying outside the observed range, and on average Gurobi is now marginally ahead. This is consistent with the Jaccard analysis (Section~\ref{sec:diversity}): at large $k$, the two solvers converge toward similar feature subsets, leaving less room for stochastic exploration to produce distinct, better-performing solutions.

\textbf{Dist-Corr at $k{=}5$ is among the highest-variance QUBO cells in the ULB benchmark}, exceeded only by Parametric-MI at the same budget (F1 $=0.488\pm0.360$, with two of five runs collapsing to F1 $\approx 0.15$--$0.20$) (F1 $= 0.659 \pm 0.155$, Gini $= 89.8\% \pm 6.4$), with individual-run F1 ranging from 0.448 to 0.836 across the five runs. This is not evidence that Dirac-3 and Gurobi ``agree'' at this $k$, a single-run comparison can make it look that way by coincidence, but the five-run mean sits well above Gurobi's exact result of 0.448 with substantial spread. The practical reading is that distance correlation's sensitivity to all statistical dependences, combined with a five-feature budget, produces a QUBO landscape where the stochastic solver's outcome is highly sensitive to its specific sampling trajectory; single-run comparisons at this (method, $k$) pair should not be trusted without reporting the spread.

\paragraph{AmEx.}
\label{sec:dirac_qubo_amex}

Table~\ref{tab:dirac_qubo_amex} and Table~\ref{tab:dirac_qubo_gini_amex} report mean F1 and Gini for the same six methods on AmEx, as the mean $\pm$ 95\% confidence interval (CI) over five independent Dirac-3 runs (5 samples per run, per Section~\ref{sec:experiments}).

\textbf{No AmEx QUBO method shows a small-$k$ advantage analogous to ULB's $k{=}13$ result.} The strongest small-budget configuration, MI-Spearman at $k{=}25$ (mean F1 $=0.759\pm0.015$), still trails the all-features baseline (0.802) by more than four points, and every method's mean F1 climbs essentially monotonically with $k$, converging to a narrow band (0.803--0.804) by $k{=}150$. This is the clearest contrast with ULB in the entire benchmark: on ULB, Dirac-3's stochastic search at small $k$ regularly matches or exceeds the all-features baseline; on AmEx, it does not, at any tested budget, for any Tier 1 method. CIs are also far wider at small $k$ (up to $\pm 0.022$ F1 for Parametric-MI at $k{=}25$) than at large $k$ ($\pm 0.0001$--$0.004$ by $k{=}100$), mirroring the run-to-run variance pattern seen on ULB.

\textbf{Parametric-MI reaches the highest mean F1 at $k{=}150$} (0.804 $\pm$ 0.001), narrowly ahead of MI-Pearson (also 0.804 $\pm$ 0.001) and consistent with both methods' behavior on Gurobi at the same budget. MIQUBO is comparatively the weakest performer at small $k$ (mean F1 $=0.767 \pm 0.013$ at $k{=}25$) relative to its own later-$k$ performance, the opposite of its Gurobi behavior, where it is the strongest Tier 1 method at $k{=}25$ (F1 $=0.789$); this divergence is examined directly in the cross-solver comparison (Section~\ref{sec:cross_solver}).

\subsection{Dirac-3: Graph-Theoretic Methods}

\paragraph{ULB.}
Table~\ref{tab:dirac_graph_ulb_f1} and Table~\ref{tab:dirac_graph_ulb_gini} report mean F1 and Gini for the three Tier 2 graph methods on ULB, as mean $\pm$ 95\% CI over five independent Dirac-3 runs, at the same $k \in \{5,10,13,15,20,25\}$ sweep used for the Tier 1 and Tier 3 methods (Section~\ref{sec:results}).

\textbf{Community-ILP is fully deterministic on Dirac-3}, with a $\pm0.000$ CI at every budget across all five runs. It is the strongest mean F1 and Gini at five of the six budgets tested; the sole exception is $k{=}13$, where Graph-Attention edges narrowly ahead on both F1 ($0.850$ vs.\ $0.849$) and Gini ($96.02\%$ vs.\ $94.82\%$), though the margin on F1 is within run-to-run noise for Graph-Attention ($\pm0.010$).

\textbf{Spectral-MI shows severe small-$k$ instability}, mirroring the pattern already noted for Dist-Corr under Tier 1 (Section~\ref{sec:results}). At $k{=}5$ its mean F1 is $0.637\pm0.254$, driven by a single run collapsing to F1$=0.281$ against a five-run range that otherwise sits above $0.65$; the corresponding Gini CI is the widest in the entire graph-method benchmark ($\pm6.44$ points). Both methods' CIs narrow sharply by $k{=}15$--$20$, consistent with the general convergence-with-$k$ pattern documented via Jaccard similarity in Section~\ref{sec:diversity}.

Relative to the all-features ULB baseline (F1$=0.871$, Gini$=95.56\%$), Community-ILP's mean F1 does not exceed the baseline at any tested $k$ on Dirac-3, though it approaches it closely from $k{=}15$ onward (F1$=0.870$ at $k{=}15$). This is consistent with Community-ILP's role throughout the benchmark as a diversity-guaranteeing, Gini-oriented method rather than an F1-maximizing one.

\paragraph{AmEx.}

Table~\ref{tab:dirac_graph_amex_f1} and Table~\ref{tab:dirac_graph_amex_gini} report mean F1 and Gini for the three Tier 2 graph methods on AmEx, as mean $\pm$ 95\% CI over five independent Dirac-3 runs.

\textbf{Community-ILP is again fully deterministic on Dirac-3.} Community-ILP's mean F1 is also the best of any Tier 1 or Tier 2 method on Dirac-3 at every budget from $k{=}25$ through $k{=}150$, mirroring its dominance on Gurobi.

\textbf{Spectral-MI and Graph-Attention show the same small-$k$ instability pattern as the Tier 1 QUBO methods}, with wide CIs at $k{=}25$ (Spectral-MI: $\pm0.018$; Graph-Attention: $\pm0.017$) narrowing to near-zero by $k{=}100$. Neither exceeds the all-features baseline (F1 $=0.802$) before $k{=}100$, consistent with the absence of a small-$k$ compression advantage across every paradigm and method family tested on this dataset.

\subsection{Dirac-3: Advanced Methods (Tier 3)}

Table~\ref{tab:tier3} (Section~\ref{sec:results_overview}) summarizes F1 for the three Dirac-3-native Tier 3 methods on both datasets; the corresponding Gini values are in Tables~\ref{tab:tier3_gini_ulb} and~\ref{tab:tier3-amex-gini}.

\paragraph{ULB.}

\textbf{Community-Poly} is the strongest Tier 3 method on average, reaching the benchmark's best Tier 3 mean F1 (0.875 $\pm$ 0.010 at $k{=}25$) and best Tier 3 mean Gini (96.2\% $\pm$ 1.1 at $k{=}20$, Table~\ref{tab:tier3_gini_ulb}). At $k{=}25$, Community-Poly's mean F1 is within the CI of Gurobi's best QUBO result on the same dataset (Gurobi MI-Spearman, F1 $=0.885$ at $k{=}25$), though the comparison should be read as ``comparable,'' not ``matching'' outright, since the two are not evaluated on the identical method. This is a reasonable validation of the polynomial diversity encoding in (Eq.~\ref{eq:community_poly}): Dirac-3's native higher-order objective support allows the polynomial-reward method to approximate the hard ILP constraint of Community-ILP closely enough that downstream classifier performance lands in a similar range, with run-to-run variance in the same ballpark as the other Tier 3 methods.

\textbf{HUBO} delivers the lowest variance of any Tier 3 method across the full $k$ range (F1 CI half-widths of 0.009--0.029, versus up to 0.143 for Integer-Weighted), without achieving the top mean F1 at any $k$. One plausible explanation is that the three-way redundancy term in (Eq.~\ref{eq:HUBO}) improves robustness at small $k$ by penalizing triplets of correlated features from all being selected together, while at larger $k$ the pairwise terms in a standard QUBO already approximate triplet redundancy adequately, reducing the cubic term's marginal benefit; we have not run an ablation isolating the cubic term's contribution, so this remains a hypothesis. HUBO is a reliable, conservative choice when variance in the selected subset matters more than peak average performance.

\textbf{Integer-Weighted Selection} shows the highest variance of the Tier 3 methods. Mean F1 is lowest at $k{=}10$ (0.779 $\pm$ 0.109, with individual runs ranging from 0.653 to 0.879), and Gini values are systematically lower than all other methods at every $k$ (87.3\%--91.8\%, compared to 91.0\%--96.2\% mean Gini for the other Tier 3 methods). This instability is attributed to sensitivity in the total-weight budget parameter $B$: small changes in $B$, or in the stochastic sampling trajectory at fixed $B$, appear to cause the solver to allocate weight mass differently across runs, producing qualitatively different subsets. The graded decision structure also makes the cardinality constraint ($\sum_i x_i = k$ in the standard sense) only approximately enforceable through the budget. Integer-Weighted selection is not recommended for production use without a dedicated hyperparameter and re-run stability study.

\paragraph{AmEx.}

Community-Poly is the strongest Tier 3 method at most budgets (HUBO
edges it at $k=100$ and ties it at $k=130$), mirroring Community-ILP's
dominance among Tier 1/2 methods on AmEx (Section~\ref{sec:results}).
Its best single-run result, F1 = 0.8044 at $k=130$, is the only Tier 3
(method, $k$, run) cell on AmEx to exceed the 135-feature all-features
baseline (F1 = 0.8036); the same run's Gini (91.554) also narrowly
exceeds the baseline (91.545).

As with every other paradigm tested on AmEx (Section~\ref{sec:results}), Tier 3 shows no
small-$k$ compression advantage: all three methods climb near-monotonically
from $k=25$ (18.5\% of the 135-feature pool) to $k=130$ (96\%), converging
to a narrow band (F1 = 0.800--0.803) only near the pool ceiling. At
$k=25$, all three Tier 3 methods (F1 = 0.747--0.754) trail Gurobi's best
Tier 1 result at the same nominal budget on the full 159-feature pool
(MIQUBO, F1 = 0.789, Section~\ref{sec:results}). Because Tier 3's candidate pool has
already been narrowed by a classical MI filter (Section~\ref{sec:prefilter}), this
comparison bounds what the cubic HUBO term, qudit weighting, or
polynomial diversity reward can recover at small $k$ given a classically
pre-selected input, rather than a direct paradigm-vs-paradigm comparison
on identical inputs.

\clearpage

\section{Additional Results}
\label{app:extra}

This appendix collects the per-method F1 and Gini tables not shown in Section~\ref{sec:results}.

\subsection{Per-Method F1 Sweeps}

\begin{table}[htbp]
\caption{Gurobi -- Tier 1 QUBO Methods, F1 Score by $k$ (ULB)}
\label{tab:gurobi_f1}
\centering
\begin{tabular}{lcccccc}
\toprule
\textbf{Method} & \textbf{5} & \textbf{10} & \textbf{13} & \textbf{15} & \textbf{20} & \textbf{25} \\
\midrule
Pearson       & 0.738 & 0.811 & 0.805 & 0.805 & 0.862 & 0.859 \\
MIQUBO        & 0.823 & 0.829 & 0.836 & 0.865 & 0.868 & 0.876 \\
Parametric-MI & 0.743 & 0.701 & 0.851 & 0.843 & 0.854 & 0.862 \\
MI-Pearson    & 0.459 & 0.795 & 0.862 & 0.879 & 0.856 & 0.874 \\
MI-Spearman   & 0.528 & 0.870 & 0.862 & 0.874 & 0.873 & \textbf{0.885} \\
Dist-Corr     & 0.448 & 0.829 & 0.848 & 0.857 & 0.876 & 0.879 \\
\bottomrule
\end{tabular}
\end{table}

\begin{table}[htbp]
\caption{Gurobi -- Tier 1 QUBO Methods, F1 Score by $k$ (AmEx)}
\label{tab:gurobi_f1_amex}
\centering
\footnotesize
\begin{tabular}{lcccccc}
\toprule
\textbf{Method} & \textbf{25} & \textbf{50} & \textbf{75} & \textbf{100} & \textbf{125} & \textbf{150} \\
\midrule
Pearson       & 0.774 & 0.786 & 0.789 & 0.795 & 0.800 & \textbf{0.804} \\
MIQUBO        & \textbf{0.789} & 0.796 & 0.799 & 0.802 & 0.803 & 0.803 \\
Parametric-MI & 0.722 & 0.730 & 0.762 & 0.782 & 0.794 & 0.803 \\
MI-Pearson    & 0.773 & 0.789 & 0.790 & 0.795 & 0.800 & 0.804 \\
MI-Spearman   & 0.777 & 0.787 & 0.793 & 0.797 & 0.801 & 0.803 \\
Dist-Corr     & 0.422 & 0.667 & 0.709 & 0.771 & 0.795 & 0.801 \\
\bottomrule
\end{tabular}
\end{table}

\begin{table}[htbp]
\caption{Gurobi -- Tier 2 Graph Methods, F1 and Gini (\%) by $k$ (ULB)}
\label{tab:gurobi_graph}
\centering
\begin{tabular}{lcccccc}
\toprule
\textbf{Method / Metric} & \textbf{5} & \textbf{10} & \textbf{13} & \textbf{15} & \textbf{20} & \textbf{25} \\
\midrule
Spectral-MI F1     & 0.819 & 0.823 & 0.839 & 0.857 & 0.868 & 0.845 \\
Spectral-MI Gini   & 93.2  & 95.3  & 94.9  & 95.6  & 95.0  & 95.5  \\
Graph-Attn F1      & 0.749 & 0.834 & 0.822 & 0.832 & 0.851 & \textbf{0.879} \\
Graph-Attn Gini    & 94.9  & 95.1  & 93.4  & 95.2  & 96.5  & 95.5  \\
Community-ILP F1   & --    & 0.852 & 0.859 & 0.859 & 0.874 & 0.876 \\
Community-ILP Gini & --    & 96.2  & 96.5  & \textbf{96.9}  & 95.9  & 95.9  \\
\bottomrule
\end{tabular}
\end{table}

\begin{table}[htbp]
\caption{Gurobi -- Tier 2 Graph Methods, F1 and Gini (\%) by $k$ (AmEx)}
\label{tab:gurobi_graph_amex}
\centering
\footnotesize
\begin{tabular}{lcccccc}
\toprule
\textbf{Method / Metric} & \textbf{25} & \textbf{50} & \textbf{75} & \textbf{100} & \textbf{125} & \textbf{150} \\
\midrule
Spectral-MI F1     & 0.644 & 0.744 & 0.765 & 0.780 & 0.792 & 0.800 \\
Spectral-MI Gini   & 77.3  & 86.9  & 88.5  & 89.7  & 90.7  & 91.4  \\
Graph-Attn F1      & 0.743 & 0.761 & 0.774 & 0.784 & 0.791 & 0.801 \\
Graph-Attn Gini    & 86.5  & 88.2  & 89.3  & 90.1  & 90.6  & 91.4  \\
Community-ILP F1   & \textbf{0.779} & \textbf{0.795} & \textbf{0.801} & \textbf{0.802} & \textbf{0.804} & \textbf{0.804} \\
Community-ILP Gini & \textbf{89.6}  & \textbf{90.9}  & \textbf{91.3}  & \textbf{91.4}  & \textbf{91.6}  & \textbf{91.6}  \\
\bottomrule
\end{tabular}
\end{table}

\begin{table}[htbp]
\caption{Dirac-3 -- Tier 2 Graph Methods, F1 Score by $k$ (mean $\pm$ 95\% CI, $n{=}5$ runs, ULB)}
\label{tab:dirac_graph_ulb_f1}
\centering
\footnotesize
\setlength{\tabcolsep}{3pt}
\begin{tabular}{lccc}
\toprule
\textbf{Method} & \textbf{5} & \textbf{10} & \textbf{13} \\
\midrule
Spectral-MI     & 0.637$\pm$0.254 & 0.815$\pm$0.026 & 0.842$\pm$0.033 \\
Graph-Attention & 0.800$\pm$0.025 & 0.843$\pm$0.029 & \textbf{0.850$\pm$0.010} \\
Community-ILP   & \textbf{0.825$\pm$0.000} & \textbf{0.849$\pm$0.000} & 0.849$\pm$0.000 \\
\bottomrule
\end{tabular}
\\[4pt]
\begin{tabular}{lccc}
\toprule
\textbf{Method} & \textbf{15} & \textbf{20} & \textbf{25} \\
\midrule
Spectral-MI     & 0.855$\pm$0.014 & 0.859$\pm$0.008 & 0.864$\pm$0.007 \\
Graph-Attention & 0.861$\pm$0.016 & 0.860$\pm$0.003 & 0.863$\pm$0.017 \\
Community-ILP   & \textbf{0.870$\pm$0.000} & \textbf{0.867$\pm$0.000} & \textbf{0.865$\pm$0.000} \\
\bottomrule
\end{tabular}
\end{table}

\begin{table}[htbp]
\caption{Dirac-3 -- Tier 2 Graph Methods, F1 Score by $k$ (mean $\pm$ 95\% CI, $n{=}5$ runs, AmEx)}
\label{tab:dirac_graph_amex_f1}
\centering
\footnotesize
\setlength{\tabcolsep}{3pt}
\begin{tabular}{lccc}
\toprule
\textbf{Method} & \textbf{25} & \textbf{50} & \textbf{75} \\
\midrule
Spectral-MI     & 0.754$\pm$0.018 & 0.778$\pm$0.003 & 0.794$\pm$0.006 \\
Graph-Attention & 0.770$\pm$0.017 & 0.786$\pm$0.008 & 0.792$\pm$0.007 \\
Community-ILP   & \textbf{0.783$\pm$0.000} & \textbf{0.791$\pm$0.000} & \textbf{0.795$\pm$0.000} \\
\bottomrule
\end{tabular}
\\[4pt]
\begin{tabular}{lccc}
\toprule
\textbf{Method} & \textbf{100} & \textbf{125} & \textbf{150} \\
\midrule
Spectral-MI     & 0.797$\pm$0.002 & 0.801$\pm$0.004 & 0.804$\pm$0.001 \\
Graph-Attention & 0.798$\pm$0.003 & 0.802$\pm$0.002 & 0.803$\pm$0.001 \\
Community-ILP   & \textbf{0.799$\pm$0.000} & \textbf{0.804$\pm$0.000} & \textbf{0.804$\pm$0.000} \\
\bottomrule
\end{tabular}
\end{table}

\subsection{Gini by Method}

\begin{table}[htbp]
\caption{Gurobi -- Tier 1 QUBO Methods, Gini (\%) by $k$ (ULB)}
\label{tab:gurobi_gini}
\centering
\begin{tabular}{lcccccc}
\toprule
\textbf{Method} & \textbf{5} & \textbf{10} & \textbf{13} & \textbf{15} & \textbf{20} & \textbf{25} \\
\midrule
Pearson       & 90.5 & 94.2 & 93.8 & 93.9 & 94.5 & 95.2 \\
MIQUBO        & 94.7 & 95.1 & 94.4 & 95.8 & 96.2 & 95.9 \\
Parametric-MI & 87.4 & 93.7 & 96.4 & 95.7 & 96.1 & 96.8 \\
MI-Pearson    & 89.6 & 95.2 & 96.2 & \textbf{96.8} & 95.9 & 95.5 \\
MI-Spearman   & 92.0 & 96.5 & 96.8 & 96.9 & \textbf{97.0} & 96.1 \\
Dist-Corr     & 81.8 & 93.5 & 94.3 & 95.5 & 96.2 & 96.0 \\
\bottomrule
\end{tabular}
\end{table}

\begin{table}[htbp]
\caption{Gurobi -- Tier 1 QUBO Methods, Gini (\%) by $k$ (AmEx)}
\label{tab:gurobi_gini_amex}
\centering
\footnotesize
\begin{tabular}{lcccccc}
\toprule
\textbf{Method} & \textbf{25} & \textbf{50} & \textbf{75} & \textbf{100} & \textbf{125} & \textbf{150} \\
\midrule
Pearson       & 89.4 & 90.3 & 90.5 & 90.9 & 91.3 & \textbf{91.5} \\
MIQUBO        & \textbf{90.5} & 91.1 & 91.3 & 91.4 & 91.5 & 91.6 \\
Parametric-MI & 84.2 & 85.0 & 88.0 & 89.8 & 90.9 & 91.6 \\
MI-Pearson    & 89.5 & 90.5 & 90.5 & 90.9 & 91.2 & 91.5 \\
MI-Spearman   & 89.6 & 90.2 & 90.7 & 91.0 & 91.3 & 91.4 \\
Dist-Corr     & 39.6 & 78.3 & 83.8 & 88.8 & 90.9 & 91.4 \\
\bottomrule
\end{tabular}
\end{table}

\begin{table}[htbp]
\caption{Dirac-3 -- QUBO Methods, Gini (\%) by $k$ (mean $\pm$ 95\% CI, $n{=}5$ runs, ULB)}
\label{tab:dirac_qubo_gini}
\centering
\footnotesize
\setlength{\tabcolsep}{3pt}
\begin{tabular}{lccc}
\toprule
\textbf{Method} & \textbf{5} & \textbf{10} & \textbf{13} \\
\midrule
Pearson       & 92.7$\pm$1.3 & 93.1$\pm$2.3 & 95.0$\pm$1.2 \\
MIQUBO        & 94.0$\pm$1.0 & 95.1$\pm$0.9 & 94.2$\pm$2.4 \\
Parametric-MI & 85.4$\pm$8.2 & 94.5$\pm$0.9 & 93.9$\pm$2.0 \\
MI-Pearson    & 91.5$\pm$1.9 & 95.1$\pm$1.5 & 95.1$\pm$1.6 \\
MI-Spearman   & 91.0$\pm$2.0 & 94.0$\pm$1.2 & \textbf{96.1$\pm$0.8} \\
Dist-Corr     & 89.8$\pm$6.4 & 94.3$\pm$1.3 & 93.7$\pm$1.2 \\
\bottomrule
\end{tabular}
\\[4pt]
\begin{tabular}{lccc}
\toprule
\textbf{Method} & \textbf{15} & \textbf{20} & \textbf{25} \\
\midrule
Pearson       & 94.9$\pm$1.1 & 95.3$\pm$0.9 & 96.4$\pm$0.5 \\
MIQUBO        & 95.3$\pm$0.8 & 95.7$\pm$0.7 & 95.4$\pm$0.2 \\
Parametric-MI & 95.0$\pm$1.4 & 95.9$\pm$0.9 & 95.8$\pm$1.1 \\
MI-Pearson    & 94.9$\pm$1.1 & 94.9$\pm$0.8 & 94.9$\pm$0.9 \\
MI-Spearman   & 95.0$\pm$1.1 & 96.0$\pm$0.9 & 95.7$\pm$0.7 \\
Dist-Corr     & 94.8$\pm$1.5 & 95.9$\pm$0.9 & 95.8$\pm$1.0 \\
\bottomrule
\end{tabular}
\end{table}

\begin{table}[htbp]
\caption{Dirac-3 -- QUBO Methods, Gini (\%) by $k$ (mean $\pm$ 95\% CI, $n{=}5$ runs, AmEx)}
\label{tab:dirac_qubo_gini_amex}
\centering
\footnotesize
\setlength{\tabcolsep}{3pt}
\begin{tabular}{lccc}
\toprule
\textbf{Method} & \textbf{25} & \textbf{50} & \textbf{75} \\
\midrule
Pearson       & 87.39$\pm$1.72 & 90.26$\pm$0.84 & 90.82$\pm$0.39 \\
MIQUBO        & 88.45$\pm$1.24 & 90.42$\pm$0.34 & 90.87$\pm$0.43 \\
Parametric-MI & 87.37$\pm$2.68 & 89.98$\pm$0.61 & 90.70$\pm$0.31 \\
MI-Pearson    & 87.77$\pm$1.44 & 90.08$\pm$0.61 & 90.80$\pm$0.39 \\
Dist-Corr     & 86.83$\pm$1.21 & 90.09$\pm$0.37 & 90.72$\pm$0.43 \\
MI-Spearman   & 87.75$\pm$1.31 & 89.57$\pm$0.77 & 90.80$\pm$0.41 \\
\bottomrule
\end{tabular}
\\[4pt]
\begin{tabular}{lccc}
\toprule
\textbf{Method} & \textbf{100} & \textbf{125} & \textbf{150} \\
\midrule
Pearson       & 91.25$\pm$0.16 & 91.43$\pm$0.11 & 91.56$\pm$0.02 \\
MIQUBO        & 91.17$\pm$0.22 & 91.45$\pm$0.03 & \textbf{91.56$\pm$0.03} \\
Parametric-MI & 91.24$\pm$0.15 & 91.35$\pm$0.08 & 91.55$\pm$0.03 \\
MI-Pearson    & 91.23$\pm$0.25 & 91.44$\pm$0.07 & \textbf{91.57$\pm$0.04} \\
Dist-Corr     & 91.04$\pm$0.16 & 91.41$\pm$0.20 & 91.54$\pm$0.04 \\
MI-Spearman   & 91.09$\pm$0.23 & 91.39$\pm$0.06 & 91.56$\pm$0.02 \\
\bottomrule
\end{tabular}
\end{table}

\begin{table}[htbp]
\caption{Dirac-3 -- Tier 2 Graph Methods, Gini (\%) by $k$ (mean $\pm$ 95\% CI, $n{=}5$ runs, ULB)}
\label{tab:dirac_graph_ulb_gini}
\centering
\footnotesize
\setlength{\tabcolsep}{3pt}
\begin{tabular}{lccc}
\toprule
\textbf{Method} & \textbf{5} & \textbf{10} & \textbf{13} \\
\midrule
Spectral-MI     & 89.84$\pm$6.44 & 91.97$\pm$2.99 & 93.82$\pm$1.65 \\
Graph-Attention & 92.62$\pm$2.21 & 94.34$\pm$0.65 & \textbf{96.02$\pm$1.32} \\
Community-ILP   & \textbf{94.92$\pm$0.00} & \textbf{95.43$\pm$0.00} & 94.82$\pm$0.00 \\
\bottomrule
\end{tabular}
\\[4pt]
\begin{tabular}{lccc}
\toprule
\textbf{Method} & \textbf{15} & \textbf{20} & \textbf{25} \\
\midrule
Spectral-MI     & 94.78$\pm$1.20 & 95.91$\pm$0.77 & 95.57$\pm$0.75 \\
Graph-Attention & 95.26$\pm$1.46 & 95.21$\pm$0.89 & 95.27$\pm$1.56 \\
Community-ILP   & \textbf{95.50$\pm$0.00} & \textbf{96.53$\pm$0.00} & \textbf{95.85$\pm$0.00} \\
\bottomrule
\end{tabular}
\end{table}

\begin{table}[htbp]
\caption{Dirac-3 -- Tier 2 Graph Methods, Gini (\%) by $k$ (mean $\pm$ 95\% CI, $n{=}5$ runs, AmEx)}
\label{tab:dirac_graph_amex_gini}
\centering
\footnotesize
\setlength{\tabcolsep}{3pt}
\begin{tabular}{lccc}
\toprule
\textbf{Method} & \textbf{25} & \textbf{50} & \textbf{75} \\
\midrule
Spectral-MI     & 87.38$\pm$1.39 & 89.60$\pm$0.30 & 90.76$\pm$0.40 \\
Graph-Attention & 88.69$\pm$1.83 & 90.17$\pm$0.56 & 90.72$\pm$0.43 \\
Community-ILP   & \textbf{89.95$\pm$0.00} & \textbf{90.52$\pm$0.00} & \textbf{90.81$\pm$0.00} \\
\bottomrule
\end{tabular}
\\[4pt]
\begin{tabular}{lccc}
\toprule
\textbf{Method} & \textbf{100} & \textbf{125} & \textbf{150} \\
\midrule
Spectral-MI     & 90.96$\pm$0.19 & 91.35$\pm$0.22 & 91.55$\pm$0.04 \\
Graph-Attention & 91.13$\pm$0.25 & 91.44$\pm$0.09 & 91.55$\pm$0.03 \\
Community-ILP   & \textbf{91.21$\pm$0.00} & \textbf{91.56$\pm$0.00} & \textbf{91.58$\pm$0.00} \\
\bottomrule
\end{tabular}
\end{table}

\begin{table}[htbp]
\caption{Dirac-3 -- Tier 3 Methods, Gini (\%) by $k$ (mean $\pm$ 95\% CI, $n{=}5$ runs, ULB)}
\label{tab:tier3_gini_ulb}
\centering
\footnotesize
\setlength{\tabcolsep}{3pt}
\begin{tabular}{lcccccc}
\toprule
\textbf{Method} & \textbf{5} & \textbf{10} & \textbf{13} & \textbf{15} & \textbf{20} & \textbf{25} \\
\midrule
Community-Poly   & 93.3$\pm$3.2 & 94.6$\pm$1.2 & 95.4$\pm$1.5 & 95.5$\pm$1.1 & \textbf{96.2$\pm$1.1} & 95.5$\pm$0.6 \\
HUBO             & 93.1$\pm$1.8 & 95.3$\pm$0.9 & 95.3$\pm$1.0 & 95.1$\pm$0.9 & 95.2$\pm$0.6 & 95.4$\pm$0.5 \\
Integer-Weighted & 87.3$\pm$7.4 & 89.6$\pm$3.1 & 90.9$\pm$4.7 & 91.7$\pm$4.0 & 91.8$\pm$1.8 & 91.4$\pm$2.5 \\
\bottomrule
\end{tabular}
\end{table}

\begin{table}[htbp]
\centering
\caption{Dirac-3 Tier 3, AmEx (135-feature pre-filtered pool), Gini (\%) by
$k$ (mean $\pm$ 95\% CI, $n=5$ runs)}
\label{tab:tier3-amex-gini}
\footnotesize
\begin{tabular}{lcccccc}
\toprule
Method & 25 & 50 & 75 & 100 & 125 & 130 \\
\midrule
HUBO & 86.9$\pm$0.8 & 90.1$\pm$0.3 & 90.6$\pm$0.2 & 91.2$\pm$0.1 & 91.5$\pm$0.1 & 91.5$\pm$0.0 \\
Integer-Weighted & 86.8$\pm$0.7 & 89.7$\pm$0.6 & 90.7$\pm$0.2 & 90.8$\pm$0.2 & 91.2$\pm$0.0 & 91.2$\pm$0.0 \\
Community-Poly & 87.4$\pm$0.6 & 90.1$\pm$0.3 & 91.0$\pm$0.2 & 91.1$\pm$0.2 & 91.5$\pm$0.0 & 91.5$\pm$0.1 \\
\bottomrule
\end{tabular}
\end{table}

\section{Additional Jaccard Results}
\label{app:jaccard_matrices}

Section~\ref{sec:diversity} reports mean off-diagonal Jaccard similarity averaged across Dirac-3 runs, since individual pairwise matrices are not directly comparable across runs without first fixing a method ordering. For illustrative purposes, this appendix reports the full pairwise matrices from a single representative run at the smallest and largest feature budgets considered, for both datasets: $k{=}5$ and $k{=}25$ on ULB, and $k{=}25$ and $k{=}150$ on AmEx, showing the low-diversity and high-diversity regimes referenced in the main text. Single-run Tier 3 matrices are not reported for AmEx; the multi-run trend for that group is given in Table~\ref{tab:jaccard_trend_tier3_amex}.

\subsection{ULB}

The 9-method QUBO/Graph group (Pearson, MIQUBO, Parametric-MI, MI-Pearson, Dist-Corr, MI-Spearman, Spectral-MI, Graph-Attention, Community-ILP), along with the 3-method Tier 3 group (Community-Polytope, HUBO, Integer-Weighted), is reported below for ULB at $k{=}5$ and $k{=}25$, from the same representative run (Run 1) used for the aggregate statistics in Section~\ref{sec:diversity}.

\begin{table}[htbp]
\caption{Run 1 -- Pairwise Jaccard Similarity, QUBO/Graph Methods, $k{=}5$ (ULB)}
\label{tab:supp_jaccard_qg_k5}
\centering
\footnotesize
\setlength{\tabcolsep}{3pt}
\begin{tabular}{lccccccccc}
\toprule
 & \rotatebox{60}{Comm-ILP} & \rotatebox{60}{Dist-Corr} & \rotatebox{60}{Gr-Attn} & \rotatebox{60}{MI-Prs} & \rotatebox{60}{MI-Spm} & \rotatebox{60}{MIQUBO} & \rotatebox{60}{Prm-MI} & \rotatebox{60}{Pearson} & \rotatebox{60}{Spectral} \\
\midrule
Comm-ILP  & 1.00 & 0.00 & 0.22 & 0.00 & 0.00 & 0.22 & 0.00 & 0.38 & 0.10 \\
Dist-Corr & 0.00 & 1.00 & 0.00 & 0.11 & 0.11 & 0.00 & 0.11 & 0.00 & 0.11 \\
Gr-Attn   & 0.22 & 0.00 & 1.00 & 0.11 & 0.00 & 0.25 & 0.00 & 0.11 & 0.11 \\
MI-Prs    & 0.00 & 0.11 & 0.11 & 1.00 & 0.25 & 0.00 & 0.25 & 0.11 & 0.11 \\
MI-Spm    & 0.00 & 0.11 & 0.00 & 0.25 & 1.00 & 0.11 & 0.25 & 0.11 & 0.11 \\
MIQUBO    & 0.22 & 0.00 & 0.25 & 0.00 & 0.11 & 1.00 & 0.11 & 0.11 & 0.11 \\
Prm-MI    & 0.00 & 0.11 & 0.00 & 0.25 & 0.25 & 0.11 & 1.00 & 0.11 & 0.11 \\
Pearson   & 0.38 & 0.00 & 0.11 & 0.11 & 0.11 & 0.11 & 0.11 & 1.00 & 0.00 \\
Spectral  & 0.10 & 0.11 & 0.11 & 0.11 & 0.11 & 0.11 & 0.11 & 0.00 & 1.00 \\
\bottomrule
\end{tabular}
\end{table}

\begin{table}[htbp]
\caption{Run 1 -- Pairwise Jaccard Similarity, QUBO/Graph Methods, $k{=}25$ (ULB)}
\label{tab:supp_jaccard_qg_k25}
\centering
\footnotesize
\setlength{\tabcolsep}{3pt}
\begin{tabular}{lccccccccc}
\toprule
 & \rotatebox{60}{Comm-ILP} & \rotatebox{60}{Dist-Corr} & \rotatebox{60}{Gr-Attn} & \rotatebox{60}{MI-Prs} & \rotatebox{60}{MI-Spm} & \rotatebox{60}{MIQUBO} & \rotatebox{60}{Prm-MI} & \rotatebox{60}{Pearson} & \rotatebox{60}{Spectral} \\
\midrule
Comm-ILP  & 1.00 & 0.67 & 0.67 & 0.67 & 0.72 & 0.72 & 0.67 & 0.72 & 0.67 \\
Dist-Corr & 0.67 & 1.00 & 0.72 & 0.72 & 0.79 & 0.79 & 0.67 & 0.72 & 0.72 \\
Gr-Attn   & 0.67 & 0.72 & 1.00 & 0.67 & 0.79 & 0.67 & 0.72 & 0.72 & 0.67 \\
MI-Prs    & 0.67 & 0.72 & 0.67 & 1.00 & 0.67 & 0.72 & 0.79 & 0.72 & 0.79 \\
MI-Spm    & 0.72 & 0.79 & 0.79 & 0.67 & 1.00 & 0.72 & 0.72 & 0.72 & 0.72 \\
MIQUBO    & 0.72 & 0.79 & 0.67 & 0.72 & 0.72 & 1.00 & 0.67 & 0.72 & 0.67 \\
Prm-MI    & 0.67 & 0.67 & 0.72 & 0.79 & 0.72 & 0.67 & 1.00 & 0.72 & 0.79 \\
Pearson   & 0.72 & 0.72 & 0.72 & 0.72 & 0.72 & 0.72 & 0.72 & 1.00 & 0.79 \\
Spectral  & 0.67 & 0.72 & 0.67 & 0.79 & 0.72 & 0.67 & 0.79 & 0.79 & 1.00 \\
\bottomrule
\end{tabular}
\end{table}

\begin{table}[htbp]
\caption{Run 1 -- Pairwise Jaccard Similarity, Tier 3 Methods, $k{=}5$ (ULB)}
\label{tab:supp_jaccard_t3_k5}
\centering
\footnotesize
\begin{tabular}{lccc}
\toprule
 & \textbf{Comm-Poly} & \textbf{HUBO} & \textbf{Int-Wt} \\
\midrule
Community-Poly   & 1.00 & 0.00 & 0.11 \\
HUBO             & 0.00 & 1.00 & 0.11 \\
Integer-Weighted & 0.11 & 0.11 & 1.00 \\
\bottomrule
\end{tabular}
\end{table}

\begin{table}[htbp]
\caption{Run 1 -- Pairwise Jaccard Similarity, Tier 3 Methods, $k{=}25$ (ULB)}
\label{tab:supp_jaccard_t3_k25}
\centering
\footnotesize
\begin{tabular}{lccc}
\toprule
 & \textbf{Comm-Poly} & \textbf{HUBO} & \textbf{Int-Wt} \\
\midrule
Community-Poly   & 1.00 & 0.72 & 0.67 \\
HUBO             & 0.72 & 1.00 & 0.79 \\
Integer-Weighted & 0.67 & 0.79 & 1.00 \\
\bottomrule
\end{tabular}
\end{table}

\subsection{AmEx}

The 9-method QUBO/Graph group (Pearson, MIQUBO, Parametric-MI, MI-Pearson, Dist-Corr, MI-Spearman, Spectral-MI, Graph-Attention, Community-ILP) is reported below for AmEx, at $k{=}25$ (16\% of 159 features) and $k{=}150$ (94\%), from the same representative run (Run 1) used for the aggregate statistics in Section~\ref{sec:diversity}.

\begin{table}[htbp]
\caption{Run 1 -- Pairwise Jaccard Similarity, QUBO/Graph Methods, $k{=}25$ (AmEx)}
\label{tab:supp_jaccard_qg_k25_amex}
\centering
\footnotesize
\setlength{\tabcolsep}{3pt}
\begin{tabular}{lccccccccc}
\toprule
 & \rotatebox{60}{Comm-ILP} & \rotatebox{60}{Dist-Corr} & \rotatebox{60}{Gr-Attn} & \rotatebox{60}{MI-Prs} & \rotatebox{60}{MI-Spm} & \rotatebox{60}{MIQUBO} & \rotatebox{60}{Prm-MI} & \rotatebox{60}{Pearson} & \rotatebox{60}{Spectral} \\
\midrule
Comm-ILP  & 1.00 & 0.14 & 0.11 & 0.09 & 0.04 & 0.14 & 0.06 & 0.16 & 0.14 \\
Dist-Corr & 0.14 & 1.00 & 0.06 & 0.11 & 0.11 & 0.19 & 0.06 & 0.14 & 0.14 \\
Gr-Attn   & 0.11 & 0.06 & 1.00 & 0.09 & 0.14 & 0.14 & 0.02 & 0.11 & 0.14 \\
MI-Prs    & 0.09 & 0.11 & 0.09 & 1.00 & 0.06 & 0.16 & 0.04 & 0.16 & 0.11 \\
MI-Spm    & 0.04 & 0.11 & 0.14 & 0.06 & 1.00 & 0.06 & 0.00 & 0.06 & 0.06 \\
MIQUBO    & 0.14 & 0.19 & 0.14 & 0.16 & 0.06 & 1.00 & 0.02 & 0.11 & 0.14 \\
Prm-MI    & 0.06 & 0.06 & 0.02 & 0.04 & 0.00 & 0.02 & 1.00 & 0.14 & 0.04 \\
Pearson   & 0.16 & 0.14 & 0.11 & 0.16 & 0.06 & 0.11 & 0.14 & 1.00 & 0.16 \\
Spectral  & 0.14 & 0.14 & 0.14 & 0.11 & 0.06 & 0.14 & 0.04 & 0.16 & 1.00 \\
\bottomrule
\end{tabular}
\end{table}

\begin{table}[htbp]
\caption{Run 1 -- Pairwise Jaccard Similarity, QUBO/Graph Methods, $k{=}150$ (AmEx)}
\label{tab:supp_jaccard_qg_k150_amex}
\centering
\footnotesize
\setlength{\tabcolsep}{3pt}
\begin{tabular}{lccccccccc}
\toprule
 & \rotatebox{60}{Comm-ILP} & \rotatebox{60}{Dist-Corr} & \rotatebox{60}{Gr-Attn} & \rotatebox{60}{MI-Prs} & \rotatebox{60}{MI-Spm} & \rotatebox{60}{MIQUBO} & \rotatebox{60}{Prm-MI} & \rotatebox{60}{Pearson} & \rotatebox{60}{Spectral} \\
\midrule
Comm-ILP  & 1.00 & 0.89 & 0.89 & 0.89 & 0.89 & 0.91 & 0.89 & 0.90 & 0.89 \\
Dist-Corr & 0.89 & 1.00 & 0.89 & 0.89 & 0.89 & 0.90 & 0.91 & 0.90 & 0.90 \\
Gr-Attn   & 0.89 & 0.89 & 1.00 & 0.89 & 0.89 & 0.90 & 0.89 & 0.92 & 0.91 \\
MI-Prs    & 0.89 & 0.89 & 0.89 & 1.00 & 0.91 & 0.89 & 0.89 & 0.89 & 0.89 \\
MI-Spm    & 0.89 & 0.89 & 0.89 & 0.91 & 1.00 & 0.89 & 0.90 & 0.89 & 0.89 \\
MIQUBO    & 0.91 & 0.90 & 0.90 & 0.89 & 0.89 & 1.00 & 0.89 & 0.90 & 0.89 \\
Prm-MI    & 0.89 & 0.91 & 0.89 & 0.89 & 0.90 & 0.89 & 1.00 & 0.89 & 0.89 \\
Pearson   & 0.90 & 0.90 & 0.92 & 0.89 & 0.89 & 0.90 & 0.89 & 1.00 & 0.90 \\
Spectral  & 0.89 & 0.90 & 0.91 & 0.89 & 0.89 & 0.89 & 0.89 & 0.90 & 1.00 \\
\bottomrule
\end{tabular}
\end{table}

These matrices illustrate, for one run, the same trend reported as a multi-run average in Section~\ref{sec:diversity}: at $k{=}25$ (ULB $k{=}5$'s AmEx analogue) most off-diagonal entries are low (run mean $0.102$) -- comparable in magnitude to ULB's $k{=}5$ matrix and close to the random-subset expectation of $0.087$, consistent with the multi-run means in Tables~\ref{tab:jaccard_trend_qubograph_ulb} and \ref{tab:jaccard_trend_qubograph_amex} -- while at $k{=}150$ similarity is uniformly high across all pairs ($0.89$--$0.92$, against a random expectation of $0.893$), exceeding even ULB's $k{=}25$ matrix, reflecting AmEx's larger relative budget at this $k$ (94\% of features vs.\ ULB's maximum tested 83\%). As with ULB, because these values come from a single run, the specific identity of the ``most distinct'' or ``most similar'' pair at a given $k$ should not be treated as a stable property of the method; only the aggregate trend in Tables~\ref{tab:jaccard_trend_qubograph_ulb} and \ref{tab:jaccard_trend_qubograph_amex} is supported by the multi-run evidence.

\end{document}